\documentclass[]{aa}
\usepackage{graphicx}
\usepackage{txfonts}
\usepackage[colorlinks]{hyperref}
\hypersetup{colorlinks=true,linkcolor=blue,citecolor=blue,filecolor=blue,urlcolor=blue,}
\usepackage{amsmath}
\usepackage{algorithm}
\usepackage{amsfonts}
\usepackage{mathrsfs}
\usepackage{bm}
\usepackage{rotating}
\usepackage{color}
\usepackage{graphicx}
\usepackage{subfigure}
\usepackage{savesym}
\usepackage[flushleft]{threeparttable}
\savesymbol{tablenum}
\restoresymbol{SIX}{tablenum}
\def \a3 {A$^3$COSMOS }
\def \ad {A$^3$C20 }

\begin{document}

   \title{A$^3$COSMOS: the infrared luminosity function and dust-obscured star formation rate density at $0.5<z<6$}

   \author{A. Traina \inst{1,2}, C. Gruppioni \inst{1}, I. Delvecchio \inst{3}, F. Calura \inst{1}, L. Bisigello \inst{4,5}, A. Feltre \inst{6}, B. Magnelli \inst{7}, E. Schinnerer \inst{8}, D. Liu \inst{9}, S. Adscheid \inst{10}, M. Behiri \inst{11,12}, F. Gentile \inst{1,2}, F. Pozzi \inst{2}, M. Talia \inst{1,2}, G. Zamorani \inst{1}, H. Algera \inst{13,14}, S. Gillman \inst{15,16}, E. Lambrides \inst{17}, M. Symeonidis \inst{18}
          }

   \institute{Istituto Nazionale di Astrofisica (INAF) - Osservatorio di Astrofisica e Scienza dello Spazio (OAS), via Gobetti 101, I-40129 Bologna, Italy
         \and
             Dipartimento di Fisica e Astronomia (DIFA), Università di Bologna, via Gobetti 93/2, I-40129 Bologna, Italy
         \and
             INAF - Osservatorio Astronomico di Brera 28, 20121, Milano, Italy and Via Bianchi 46, 23807 Merate, Italy
         \and
             INAF - Osservatorio Astronomico di Padova, Vicolo dell'Osservatorio, 3, I-35122, Padova, Italy
         \and
             Dipartimento di Fisica e Astronomia, Università di Padova, Vicolo dell'Osservatorio, 3, I-35122, Padova, Italy
         \and 
             INAF - Osservatorio Astrofisico di Arcetri, Largo E. Fermi 5, 50125, Firenze, Italy
         \and 
             Université Paris-Saclay, Université Paris  Cité, CEA, CNRS, AIM, F-91191 Gif-sur-Yvette, France
         \and
             Max Planck Institut für Astronomie, Königstuhl 17, D-69117 Heidelberg, Germany
         \and 
             Max-Planck-Institut für extraterrestrische Physik, Gießenbachstraße 1, 85748, Garching b. München, Germany
         \and 
             Argelander-Institut für Astronomie, Universität Bonn, Auf dem Hügel 71, 53121 Bonn, Germany
         \and
             SISSA, Via Bonomea 265, 34136 Trieste, Italy
         \and 
             IRA-INAF, Via Gobetti 101, 40129 Bologna, Italy
         \and
             Hiroshima Astrophysical Science Center, Hiroshima University, 1-3-1 Kagamiyama-Higashi, Hiroshima 739-8526, Japan
         \and    
             National Astronomical Observatory of Japan, 2-21-1 Osawa, Mitaka, 169-8555, Tokyo, Japan
         \and 
             Cosmic Dawn Center (DAWN)
         \and
             DTU-Space, Technical University of Denmark, Elektrovej 327, DK-2800 Kgs. Lyngby, Denmark
         %\and
             %The William H. Miller III Department of Physics \& Astronomy, Johns Hopkins University, Baltimore, MD 21218, USA
         \and 
             NPP Fellow, NASA Goddard Space Flight Center, Greenbelt, MD 20771, USA             
         \and
             Mullard Space Science Laboratory, University College London, Holmbury St. Mary, Dorking, Surrey RH5 6NT, UK
             }

   \date{Received ??; accepted ??}

% \abstract{}{}{}{}{} 
% 5 {} token are mandatory
 
  \abstract
   {}
   {We leverage the largest available Atacama Large Millimetre/submillimetre Array (ALMA) survey from the archive (A$^3$COSMOS) to study to study infrared luminosity function and dust-obscured star formation rate density of sub-millimeter/millimeter (sub-mm/mm) galaxies from $z=0.5\,-\,6$.}
 {The \a3 survey utilizes all publicly available ALMA data in the COSMOS field, therefore having inhomogeneous coverage in terms of observing wavelength and depth. In order to derive the luminosity functions and star formation rate densities, we apply a newly developed method that corrects the statistics of an inhomogeously sampled survey of individual pointings to those representing an unbiased blind survey.}
  % results
    { We find our sample to mostly consist of massive ($M_{\star} \sim 10^{10} - 10^{12}$ $\rm M_{\odot}$), IR-bright ($L_* \sim 10^{11}-10^{13.5} \rm L_{\odot}$), highly star-forming (SFR $\sim 100-1000$ $\rm M_{\odot}$ $\rm yr^{-1}$) galaxies. We find an evolutionary trend in the typical density ($\Phi^*$) and luminosity ($L^*$) of the galaxy population, which decrease and increase with redshift, respectively. Our IR LF is in agreement with previous literature results and we are able to extend to high redshift {($z > 3$)} the constraints on the knee and bright-end of the LF, derived by using the Herschel data. Finally, we obtain the SFRD up to $z\sim 6$ by integrating the IR LF, finding a broad peak from $z \sim 1$ to $z \sim 3$ and a decline towards higher redshifts, in agreement with recent IR/mm-based studies, within the uncertainties, thus implying the presence of larger quantities of dust than what is expected by optical/UV studies.}  
   {}

   \keywords{galaxies: evolution – galaxies: high-redshift – galaxies: luminosity function, mass function  –
submillimeter: galaxies -  surveys}

   \titlerunning{The IR LF and SFRD at $z \sim 0.5-6$ from the A$^3$COSMOS survey}
   \authorrunning{A. Traina et al.}
   \maketitle
   
%
%-------------------------------------------------------------------

\section{Introduction}\label{sec:intro}

Understanding how galaxies formed and evolved is one of the open questions of modern astrophysics. This can be addressed in many different ways, using information across the whole electromagnetic spectrum.
One of the best approaches consists in the study of galaxy samples across a wide range of redshift and luminosity, enabling the derivation of statistical properties as a function of the redshift, such as the luminosity function (LF) and the cosmic star formation rate density (cSFRD). These quantities are fundamental to probe the statistical nature of various galaxy populations at different cosmic times, as well as to study the mass assembly process in galaxies at different epochs.
\par Up to $z \sim 2-3$ the SFRD has been well studied and accurately measured thanks to both optical-ultraviolet (UV) and infrared (IR) facilities, such as the \textit{Galaxy Evolution Explorer} (GALEX), the \textit{Hubble Space Telescope} (HST) and the \textit{Herschel Space Observatory} \citep[see, e.g.,][ for a review]{Dahlen2007sfrd,reddy2009sfrd,cucciati2012sfrd,gruppioni2013lf,magnelli2013ir, madau2014sfrd}.
These works have revealed an increase of the SFRD with redshift, which at all epoches is dominated by the obscured IR component \citep[$\sim80\%$, ][]{khusanova2021sfrd}. In particular, a key element of this IR component comes from dusty star forming galaxies (DSFGs) or sub-millimeter galaxies \citep[SMGs,][]{smail1997smg,hughes1998smg,barger1998smg,blain2002smg,casey2014review}. These objects, more common at $z \sim 2-2.5$ \citep[][]{chapman2003smg,wardlow2011smg,yun2012smg}, are characterized by large infrared (IR, $8~\mu$m $< \lambda < 1$ mm) luminosity ($>10^{12}$ $\rm L_{\odot}$), stellar masses ($>10^{10}$ $ \rm M_{\odot}$) \citep[][]{chapman2005smg,simpson2014smg}, high star formation rates (SFRs) \citep[$>100$ $\rm M_{\odot}$yr$^{-1}$,][]{magnelli2012sfr,swinbank2014sfr}, which makes them the main contributors to the SFRD at these redshifts.
\par Optical/UV based studies have made it possible to compute the SFRD up to $z \sim 7-8$ \citep[see e.g.,][]{bouwens2014sfrd, oesch2015sfrd, laporte2016sfrd, oesch2018sfrd} and even $z \sim 10$ \citep[e.g.,][]{harikane2023jwst} with the James Webb Space Telescope (JWST), extending our knowledge on the star formation activity to very early epochs of the universe. However, these studies can be potentially biased by the observing band, i.e., the rest-frame UV, which is highly affected by dust obscuration. Indeed, the dusty contribution is only retrieved from dust correction measured in the UV.
These corrections are still very uncertain, as thermalization and re-emission by dust in the IR-mm bands has been shown to be significantly contributing to the SFRD even at $z > 3-4$ \citep[see e.g.,][]{magdis2012ir,magnelli2013ir,bethermin2015ir,casey2018ir,zavala2018ir, gruppioni2020alpine,algera2022sfrdALMA}. Therefore, it is crucial to determine the contribution of galaxies selected in the IR band, where this emission is produced. While attempts have been made to constrain the SFRD$_{\rm IR}$ at $z>3$ in the past using single-dish IR-mm surveys, only the advent of ALMA opened up this possibility. The unprecedented sensitivity reached by ALMA, coupled with the assembly of unbiased samples in the mm bands \citep[][]{hodge2013aless,staguhn2014almasurvey,zavala2018ir,franco2018smg}, allows for the study of evolution of these galaxies up to higher redshifts, thus covering the $z > 3$ range still affected by many uncertainties (poor statistics, bias in the IR luminosity). Recent works using sub-mm/mm samples \citep[e.g.,][]{gruppioni2020alpine, algera2022sfrdALMA} support a scenario in which the SFRD shows a plateau rather than a significant decrease at $z=2-6$. These first studies are, however, limited by statistics and larger samples are required to better constrain the SFRD at higher redshift.
\par In this perspective, the Automated mining of the ALMA Archive in COSMOS \citep[A$^3$COSMOS,][]{liu2019a32,liu2019a31}, which is a compilation of all ALMA observations in the COSMOS field, represents the largest ALMA survey to date. The fact that the survey is in the COSMOS field \citep[][]{scoville2007cosmos}, where a large wealth of multiwavelength data are available, including the COSMOS2020 catalog \citep[][]{weaver2022cosmos2020}, makes it ideal to perform statistical studies on the nature and evolution of star forming galaxies over a large range of redshifts and luminosities. However, it is not a purely blind survey, which instead is needed to perform statistical studies. Moreover, the individual pointings composing the survey are at different observing wavelengths and have different sensitivities, making it even more inhomogeneous. For these reasons, we developed, within the \a3 collaboration, a new method specifically tailored to turn a targeted survey, composed by an arbitrary number of pointings (isolated or overlapping), each with its own limiting flux (radially varying within the pointing) and observing band, into a “blind-like” (targeted unbiased) one, thus allowing the derivation of the main statistical properties of large galaxy samples over cosmic time. However, it is important to bear in mind that the conversion into a blind survey can be affected by uncertainties related to the assumptions made (see Section \ref{sec:blind}), mainly on the removal of the pointed target and/or clustered sources as well as on the conversion curves for the RMS.
In order to take advantage of the most recent A$^3$COSMOS\footnote{\href{https://sites.google.com/view/a3cosmos}{https://sites.google.com/view/a3cosmos}} and multiwavelength \citep[COSMOS2020][]{weaver2022cosmos2020} catalogs, we have performed a new catalog match and SED-fitting analysis using the \textsc{python} “Code Investigating GALaxy Emission" \citep[\texttt{CIGALE};][]{boquien2019cigale} SED fitting tool. We then derived the IR LF and we present new estimates for the dust-obscured SFRD from $z \sim 0.5$ to $z \sim 6$. The paper is organized as follows: in Section \ref{sec:sample} we present in detail the \a3 survey and the multiwavelength sample, as well as the available photometry; in Section \ref{sec:cigale} we describe the SED fitting and the results obtained for the physical properties; in Section \ref{sec:blind} we illustrate our new method to turn the \a3 survey from a pointed into a blind one; in Section \ref{sec:lf} we present our estimates of the IR LF; in Section \ref{sec:sfrd} we derive the SFRD; finally, in Section \ref{discussion} we report a summary of our conclusions. 
\par Throughout the present work, we assume a \cite{chabrier2003imf} stellar
initial mass function (IMF) and adopt a $\Lambda$CDM cosmology with $H_{0} = 70$ $\rm km$ $\rm s^{-1}$ $\rm Mpc^{-1}$, $\Omega_{\rm m} = 0.3$, and $\Omega_{\Lambda} = 0.7$.
%= 70 km s$^{−1}$ Mpc$^{−1}$, $\Omega_{m} = 0.3$, and $\Omega_{\Lambda} = 0.7$.

%--------------------------------------------------------------------
%%%%%%%%%%%%%%%%%%%%%%%%%%%%%%%%
%
%Sample description
%
%%%%%%%%%%%%%%%%%%%%%%%%%%%%%%%%
\section{Catalog descriptions}\label{sec:sample} 

In this section we present the catalogs used in this work and describe their main features.
We provide a brief description of the automated pipeline used to obtain the catalogue. A more detailed description of how the pipeline was developed can be found in \cite{liu2019a31} and \textcolor{blue}{Adscheid et al., prep}; here we briefly describe it.
The survey is built by downloading the ALMA pointings from the archive with automatic pipelines, using the Python package astroquery \citep[][]{ginsburg2019astroquery}. For calibration and creation of continuum images from the raw data, the Common Astronomy Software Application \citep[CASA;][]{mccullin2007casa} was used. Sources were extracted blindly and in prior mode, using Python Blob Detector and Source Finder \citep[PyBDSF;][blind]{mohan2015pybdsf} and GALFIT \citep[][ prior]{peng2002galfit,peng2010galfit}, utilizing prior source positions from multi-wavelength catalogs covering the COSMOS field. The matching with the priors was done with a $1"$ radius to reduce multiple associations. Finally, to limit the number of spurious sources, a minimum peak signal-to-noise threshold of 4.35 $\sigma$ (for the prior) was applied, resulting in a global spurious fraction of less than $12\%$ \citep[see Figure 8 from ][]{liu2019a31}.
%In particular, focusing on the advantages offered by the combination of the wide photometric coverage of the COSMOS2020 with the ALMA archival observations in the study of large sample of galaxies.

\subsection{\a3 catalog}
%Since the first light of the ALMA interferometer, a large amount of observations has been collected. %Among these, a large number of ALMA observations are dedicated to high-$z$ ($z>2-3$) galaxies, allowing the community to access a large sample of such objects.
The large amount of observation collected by the ALMA interferometer has already been explored in some recent works \citep[see, e.g.,][]{decunha2015aless,bouwens2016aspecs,fujimoto2017alma,scoville2017alma,zavala2018ir}, focusing on the physical properties of the high-$z$ galaxy population. However, in order to investigate galaxy properties (such as the gas and dust content) of statistically significant samples, a systematic mining of the archive is needed. In this context, the \a3 project \citep[][]{liu2019a31,liu2019a32} aims at building a catalog of galaxies from ALMA archival images, processed homogeneously, in the COSMOS field. In this way, it is possible to retrieve both targeted and serendipitously detected objects in the individual pointings, building a statistically robust catalog of sub-mm detected sources.
Within the \a3 survey, two different catalogs are available. The first contains sources blindly extracted from the images, while the second one is a prior-based catalog, using optical/near-infrared positional priors (see Section \ref{subsec:cosmos2020}). In this work, we used the prior version of the catalog (1620 sources), since it allows us to construct the broad-band (from UV up to sub-mm/mm) spectral energy distribution (SED) of our sample.

\subsection{COSMOS2020 catalog}\label{subsec:cosmos2020}
The COSMOS field \citep[][]{scoville2007cosmos} is among the best studied extragalactic deep fields, owing to an unparalleled multi-wavelength photometric coverage, including X-rays \citep[][]{elvis2009xcosmos,civano2012xcosmos,civano2016xcosmos,marchesi2016xcosmos}, UV \citep[][]{zamojski2007uvcosmos}, optical \citep[][]{capak2007cosmos,Leauthaud2007ocosmos,taniguchi2007ocosmos,taniguchi2015ocosmos}, near-IR \citep[][]{mccracken2010nircosmos,mccracken2012nircosmos}, mid-IR \citep[][]{sanders2007mircosmos,lefloch2009mircosmos}, far-IR \citep[][]{lutz2011fircosmos,oliver2012fircosmos}, submillimeter \citep[][]{geach2017submmcosmos}, millimeter \citep[][]{bertoldi2007mmcosmos,Aretxaga2011mmcosmos} and radio \citep[][]{schinnerer2010radiocosmos,smolcic2017radiocosmos} bands. This has enabled the construction of large statistical samples of galaxies with measured stellar mass ($M_{\star}$) and star-formation rate (SFR) based on their photometric points via the SED-fitting technique. Over the past years, the release of several COSMOS catalogs based on different selection bands \citep[][]{capak2007cosmos,ilbert2009cosmos,ilbert2013cosmos,muzzin2013cosmos,laigle2016cosmos} has widened our possibility to investigate extended samples of galaxies spanning large ranges of physical properties. 
\par The most recent release of the COSMOS photometric catalog \citep[i.e. COSMOS2020,][]{weaver2022cosmos2020} is characterized by the addition of new data from the Hyper Suprime-Cam (HSC) Subaru Strategic Program (SSP) PDR2 \citep[][]{aihara2019hsc}, new data from the DR4 \citep[][]{moneti2023uvdr4} of the Visible Infrared Survey Telescope for Astronomy (VISTA) and all the Spitzer IRAC data in the COSMOS field.
Moreover, it contains two independently derived photometric datasets. One (\texttt{the CLASSIC}) retrieved with classical aperture photometry on PSF-homogenized images using \texttt{IRACLEAN} \citep[][]{hsieh2012irac}, and another one (\texttt{the FARMER}) derived using a PSF-fitting tool \citep[\texttt{the Tractor;}][]{lang2016tractor} to extract the 
photometry. The covered area is $\sim 1.77$ deg$^2$ and the total number of sources in \texttt{the CLASSIC} version is 1,720,700. See \cite{weaver2022cosmos2020} for a detailed description of the two methods and catalogs. In this paper we use the \texttt{CLASSIC} version of the COSMOS2020 catalog.

\subsection{Our sample}
\begin{figure}[]
\centering
{\includegraphics[width=.48\textwidth]{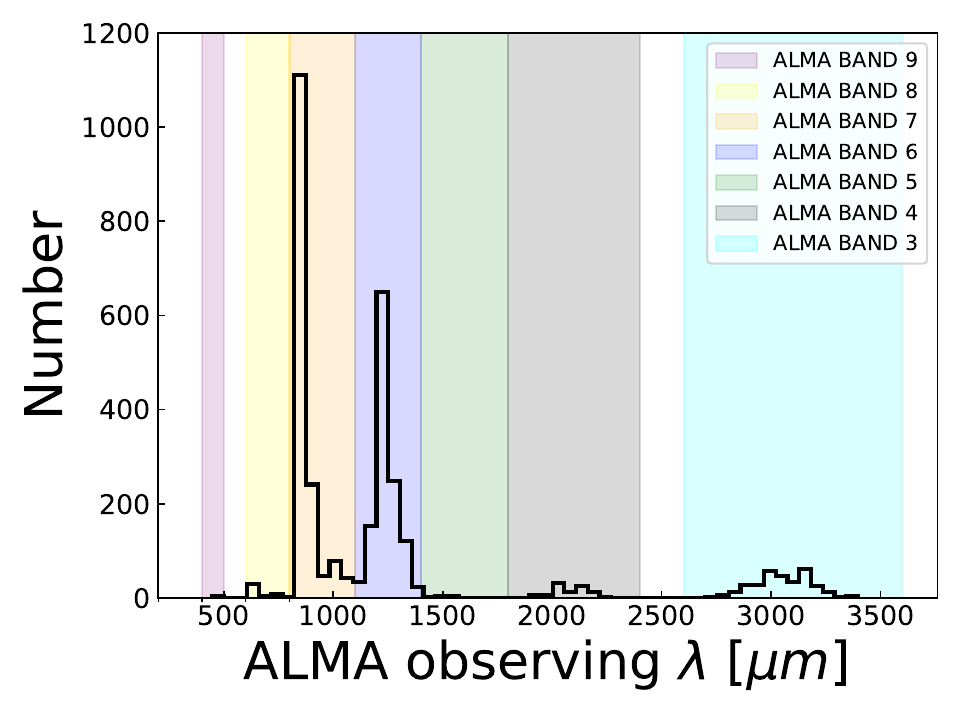}}
\caption{\small{Number of pointings as a function of observing wavelength in the \a3 database. The wavelength ranges of the ALMA bands are plotted as color-shaded regions. The most populated bands are 6 and 7, with $\sim 2000$ pointings.}}
\label{fig:a3cosmos_pointings_histo}
\end{figure}

\begin{figure*}[]
\centering
{\includegraphics[width=1.\textwidth]{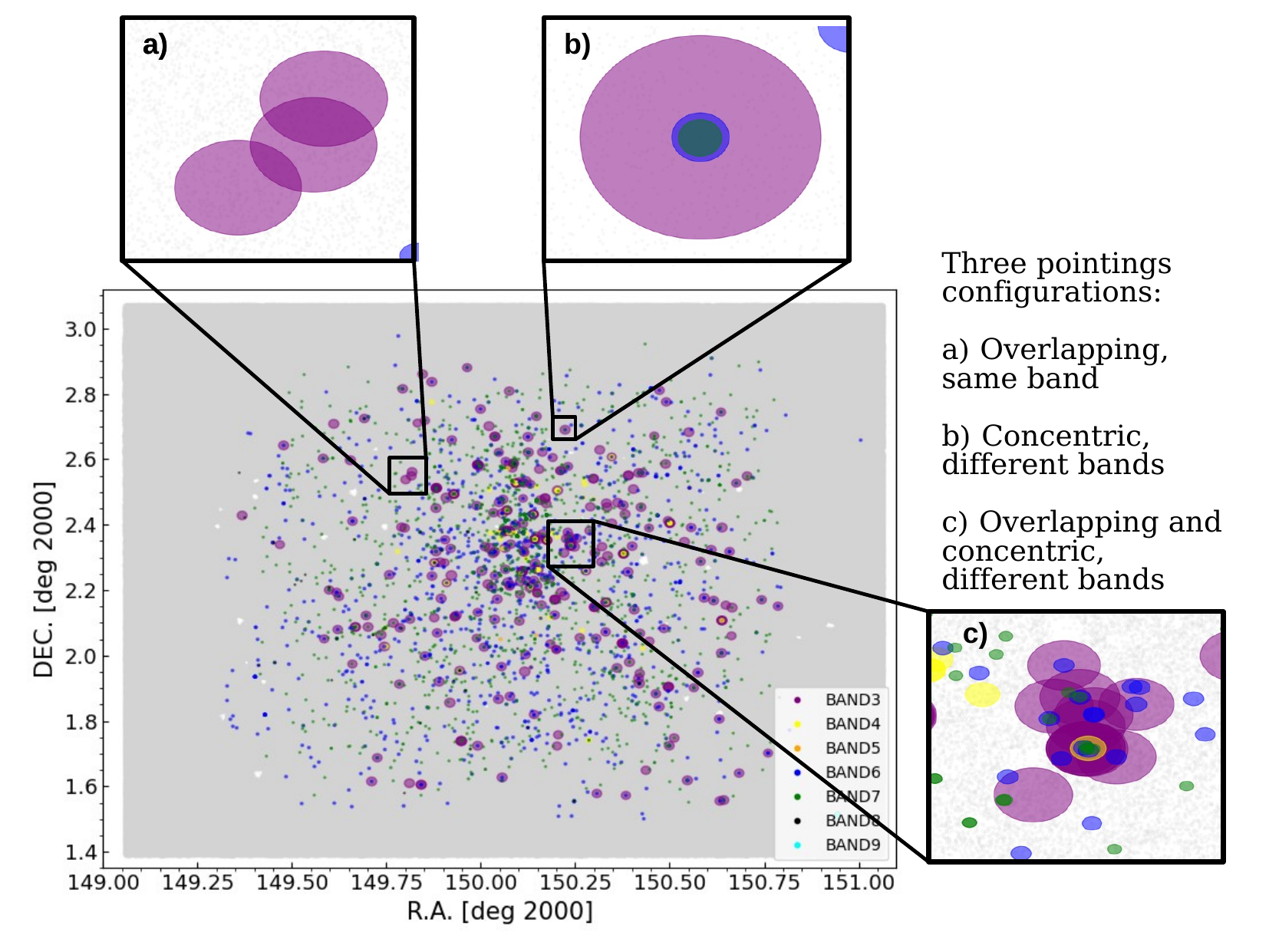}}
\caption{\small{Archival ALMA observations in the COSMOS field used in this study. Individual pointings are plotted with different colors representing the observed ALMA bands. In the background, the COSMOS field is shown in grey. We show three zoom-in regions representative of possible classes of pointing configurations. The top-left panel (a) shows three overlapping pointings in the same band; the top-right panel (b) presents three concentric pointings in different bands; in the bottom-right panel (c) overlapping and concentric pointings, in different bands, are reported. \\}}
\label{fig:a3cosmos_survey_pointings}
\end{figure*}

In this work we use the latest version from the \a3 database (\textcolor{blue}{Adscheid et al.} in prep.). %, which includes all data public by .
This version combines the already tested process of the automatic mining of the ALMA archive with the new photometry presented in the COSMOS2020 catalog (the new \a3 catalog based on prior extraction from COSMOS2020 catalog is hereafter called \ad). \ad is composed of 3215 individual pointings, coming from 171 different ALMA projects, covering ALMA bands from 3 to 9 %(see Table \ref{tab:available_pointigns}).
In Figure \ref{fig:a3cosmos_pointings_histo}, the wavelength distribution of the different pointings is reported. The ALMA bands are shown with different color-shaded areas. It is clear that the vast majority ($\sim 80 \%$) of the sample comes from ALMA bands 7 (orange area) and 6 (blue region), with more than 2500 observations available. In Figure \ref{fig:a3cosmos_survey_pointings} we show the spatial distribution of the pointings in the survey, color-coded by the observed-frame wavelength. We also highlight three different pointing configurations, as representative of their complex spatial distribution in the survey: a) a case of partially overlapping pointings in the same band; b) concentric pointings in different bands; c) an extreme case of N>10 overlapping pointings in different bands. For further details see Section \ref{sec:blind}.
\par We select galaxies above $4.35\sigma$ (with $\sigma$ being the local RMS at the position of each source, see \textcolor{blue}{Adscheid et al.}, in prep.), and it consists of 1620 sources with flux in at least one ALMA band. For our sample, $25\%$ (441/1620) of the sources have a spectroscopic redshift (spec-$z$) and for 1069/1620 we have used the photo-$z$ in the COSMOS2020 catalog. The spec-$z$ are compiled from the literature \citep[e.g., ][]{riechers2013specz,capak2015specz,smolcic2015specz,brisbin2017specz,lee2017specz}, the catalog by
M. Salvato et al. (version 2017 September 1; available internally to the COSMOS collaboration) and the ALMA archive \citep[][]{liu2019a31}. The photo-$z$ used here are from \cite{salvato2011photoz,davidzon2017photoz,delvecchio2017photoz,jin2018superdebl} and derived from either the \texttt{CLASSIC} or \texttt{FARMER} version of the COSMOS2020 catalog. Finally, 110/1620 sources do not have any redshift information. For this subsample, we computed their photo-$z$ using \texttt{CIGALE} \citep[][]{boquien2019cigale} as described in the next section.

%\begin{table}[]
%\small
%\centering
%\caption{Number of pointings available for each ALMA band.}
%\begin{tabular}{lcccccccc}
%\hline
%\hline
%\\
%\textbf{ALMA BAND}                                                                      & \textbf{3} & \textbf{4} & \textbf{5} & \textbf{6} & \textbf{7} & \textbf{8} & \textbf{9} \\ \\\hline \\
%\multicolumn{1}
%{c}{\textbf{\begin{tabular}[c]{@{}c@{}}NUMBER OF\\ POINTINGS\end{tabular}}}        &  318       &   94         &    8          & 1232       & 1521       &   40         &    2      \\    \\ \hline \hline
%\end{tabular}
%\label{tab:available_pointigns}
%\end{table}

%%%%%%%%%%%%%%%%%%%%%%%%%%%%%%%%
%
%SED fitting
%
%%%%%%%%%%%%%%%%%%%%%%%%%%%%%%%%

\section{CIGALE SED fitting}\label{sec:cigale} 
%A powerful and routinely used method to investigate the physical properties of samples of galaxies is SED fitting. %In this way, using the observed fluxes in a given number of bands, it is possible to derive information on the individual objects by fitting them with a combination of models representing the different emission components.
In this work, we decided to perform the SED fitting of the \a3 galaxies using \texttt{CIGALE}, a python SED fitting code based on the energy balance between the UV/optical emission by stars and the re-emission in the IR/mm by the dust. \texttt{CIGALE} is a highly flexible code, which allows one to choose among different individual templates for each emission components (e.g., stellar optical/UV emission, cold dust emission, AGN), across a broad parameter space. %Among the main feature of this tool, there is the possibility to compute photo-$z$, using a pre-selected redshift grid as input to derive a best-fit SED even in the absence of optical/UV data-points.
Furthermore, one of the most important features is the availability of active galactic nucleus (AGN) templates, which can be easily included in the fit, allowing a decomposition between star formation-powered and AGN-powered IR emission. In particular, derive the IR luminosity from the SED is crucial to compute the IR LF.
\par In the following sections, we report the available photometry and the individual components used to perform the SED fitting, following recent SED-based studies \citep[][]{ciesla2017sed,lofaro2017sed,malek2018sed,pearson2018sed,buat2019sed,donevski2020sed}. Moreover, when needed, we included in the fit an input grid of redshifts spanning between $z=0$ and $z=8$ (with a step of $\Delta z = 0.1$) to derive the best photo-$z$, if missing.

\subsection{Photometric coverage}
The \ad catalog, being the combination of COSMOS2020 catalog and the archival ALMA observations, takes advantage of a large photometric coverage, from the UV to the FIR/mm. To perform the SED-fitting, we have considered the following filters available in the COSMOS2020 catalog \citep{weaver2022cosmos2020}: CFHT MegaCam \textit{u}; Subaru Suprime Cam \textit{i}, \textit{B}, \textit{V}, \textit{r}, \textit{z}; Subaru HSC \textit{y}; VISTA VIRCAM \textit{Y}, \textit{J}, \textit{H}, \textit{Ks}; Spitzer IRAC channel 1, 2, 3 and 4; Spitzer MIPS 24 $\mu$m; Herschel PACS at 100 and 160 $\mu$m; Herschel SPIRE at 250, 350, 500 $\mu$m; JMCT SCUBA2 at 850 $\mu$m; ASTE AzTEC (1 mm) and IRAM MAMBO (1.2 mm). 
\par To deal with all the ALMA frequency setup present in the \a3 database, we have built artificial filters to be provided to CIGALE, corresponding to each observing wavelength in the \a3 catalog. The filters are centered at that specific wavelength and are box-like, with their width being equal to 16 GHz.%the difference between two central $\lambda$ of subsequent filters. 
With this procedure, we added up to 330 continuous ALMA filters between 446 $\mu$m and 3325 $\mu$m to the \texttt{CIGALE} database.

\subsection{CIGALE input templates}
We modeled the stellar population with the \texttt{bc03} stellar population synthesis model \citep[][]{bruzual2003ssp}, to build the SED stellar component, and a delayed star formation history (SFH) with an optional second burst of star formation. We selected the \texttt{dustatt\_modified\_CF00} \citep[][]{charlot2000dustatt} template to model the attenuation by dust and the \texttt{dl2014} \citep[][]{draine2014dustem} to model the dust emission, both based on the assumption of having two different attenuation/emission sources represented by birth clouds and diffuse ISM. 
\par Finally, we use the \texttt{fritz2006} module \citep[][]{fritz2006agnmodel,feltre2012agnmodel} to model the AGN component in the SED. The AGN emission is described with a radiative transfer model which takes into account primary emission coming from the central engine (i.e. the accretion disc), scattered emission produced by dust and a thermal component of the dust emission. The individual input parameters are described in detail in Appendix \ref{sec.foo}.

\subsection{SED fitting results}\label{subsec:sedres}
We performed the SED fitting for the 1620 sources of our sample using \texttt{CIGALE} with the modules described above. For our purposes, we obtained the following output parameters: dust luminosity (i.e., the IR 8-1000 $\mu$m luminosity), stellar mass, AGN fraction ($f_{AGN}$) contributing to the $5-40~\mu$m total emission and $\tau$ (equatorial optical depth at 9.7$\mu$m) and $\Psi$ (angle between equatorial axis and line of sight) parameters of the \texttt{fritz2006} model and redshift for 110 sources. %Where needed, we put an upper limit as a three sigma level of the flux.
\par Among the 1620 sources for which we performed the fit, we first remove those presenting a high reduced $\chi^2$ (>10). Since our goals are strictly linked to the IR emission of these sources, we also computed the ratio between the ALMA observed flux and the best-fit flux at the same wavelength; sources with a ratio greater than the $5 \sigma$ of the ratio distribution were removed and classified as bad SED, if we were not able to re-perform an acceptable fit. The $5 \sigma$ threshold has been selected to be consistent with the “good SED” selection performed by \cite{liu2019a31} and \textcolor{blue}{Adscheid et al. (prep)} in the catalog construction.
We obtained a good fit for 1411/1620 galaxies, with 43/110 sources having no initial associated redshift. The remaining 67/110 galaxies without redshift had a bad fit (i.e., the IR-mm photometry is not matching the optical part of the SED), so we decided to exclude them from the final sample, but we consider them in the incompleteness estimate. 
\par
\begin{figure*} 
\centering
\includegraphics[width=1.1\textwidth]{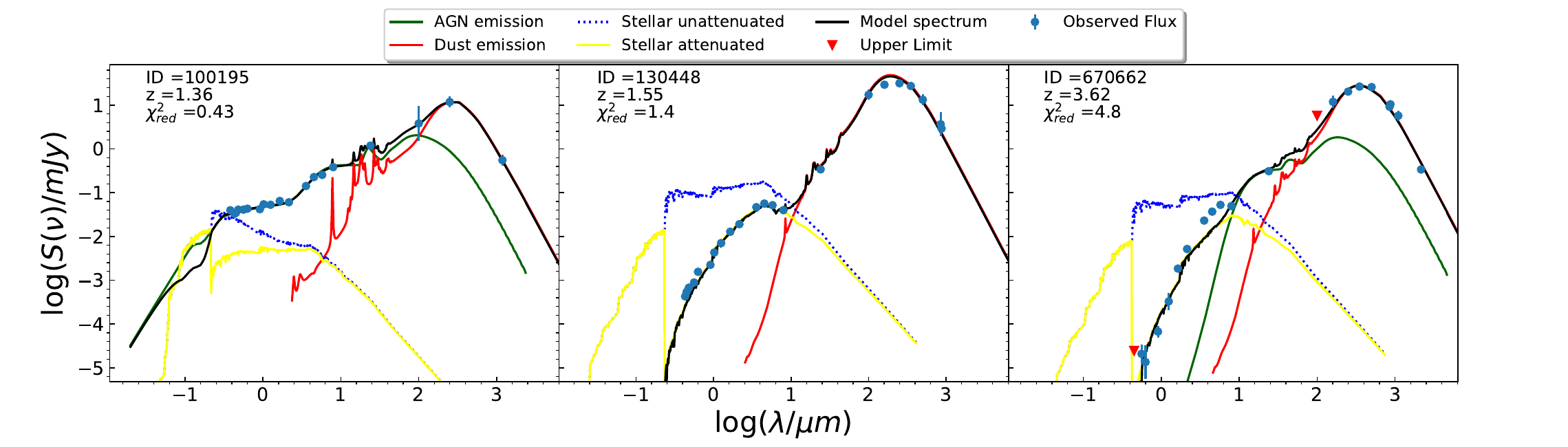}
\caption{Examples of SED-fitting results for three different classes of objects. From left to right: unobscured AGN SED, SED without an AGN contribution and obscured AGN component. The blu filled circles and the red triangles represent data points and upper limits, respectively. The best-fit model is plotted as a black solid line. The stellar attenuated and unattenuated, the dust emission and the AGN emission are reported as yellow solid line, blue dashed line, red and green solid lines, respectively. \\}
\label{fig:sed_examples}
\end{figure*}

 In Figure \ref{fig:sed_examples} we show the best-fitting SED of three representative objects: a Type II AGN, a Type I AGN and a galaxy without an AGN component.
\par
\begin{figure}[]
\centering
{\includegraphics[width=.5\textwidth]{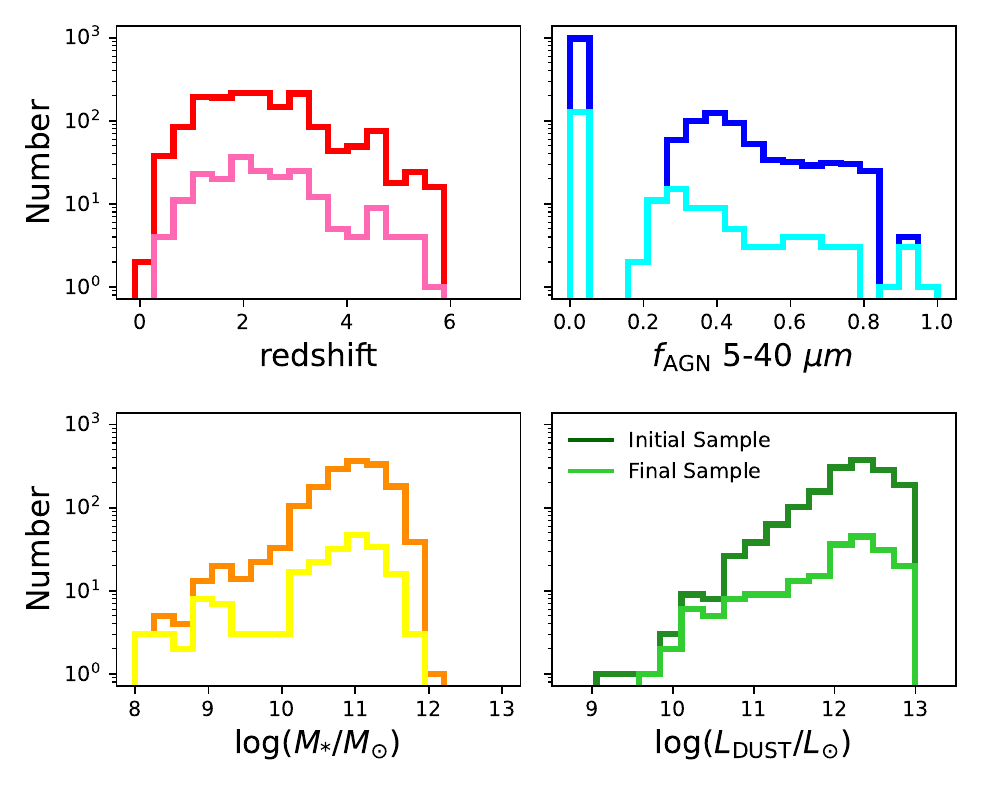}}
\caption{\small{Distributions of the main physical parameters obtained through SED fitting for the initial and final, reduced, sample. In the top-left panel the redshift distribution is shown; the top-right panel represents the AGN fraction distribution, computed in the 5-40 $\mu$m range; the bottom-left panel reports the logarithm of stellar mass distribution, in units of solar masses; finally, the bottom-right panel shows the logarithm of the dust luminosity in solar luminosities.\\}}
\label{fig:histo_sed_fitting}
\end{figure}

 In Figure \ref{fig:histo_sed_fitting} we report the redshift distribution (top left panel) of the sample, as well as the results from the SED fitting for stellar mass (bottom left), dust luminosity (bottom right) and AGN fraction (top right panel), for both the initial (1411) and final (i.e., after the cut, see Section \ref{sec:blind}, 189 galaxies). The redshift distribution peaks between $z \sim 1.5$ and $z \sim 3$.
\par The galaxies in the sample are massive with a peak in the distribution of stellar mass at log($M_{\star}/M_{\odot})  \sim 11$, consistent with them being DSFGs \citep[][]{chapman2005smg,simpson2014smg}, although our sample contains sources with masses as low as $10^{8}$ M$_{\odot}$. The \ad galaxies are on average IR-bright, most of which with IR luminosities spanning from $10^{11}$ $L_{\odot}$ to $10^{13}$ $L_{\odot}$.
Using the best-fitting templates for each source, we are able to compute the AGN luminosity in a given wavelength range and, thus, the fractional AGN contribution to the total emission in that range. The 5-40 $\mu$m range is particularly sensitive to the presence of an AGN, therefore we considered this wavelength interval to derive the fraction of AGN, $f_{AGN}$, contributing to the luminosity in this range. It can be seen that the distribution is bimodal, with most sources ($\sim 65\%$) having an AGN fraction near 0 (i.e., they are likely not hosting an AGN) and the remainder having a $f_{AGN}$ higher than $\sim 0.2$, peaking at $\sim 0.4$ with a tail up to $0.8-0.9$. The gap between $f_{AGN}=0$ and higher values is due to the grid we adopted for the SED fitting. Despite the $\sim 35\%$ of sources is including an AGN component (the mean  $f_{AGN}$ is $\sim 0.4$) in the best-fit SED, only a small fraction (\textbf{$\sim 15\%$}) is AGN-dominated (i.e., $f_{\rm AGN} > 0.5$). Also, since the AGN emission is mostly present in the $5-40$ $\mu$m range, the contribution to the $8-1000$ $\mu$m is only a small fraction. For consistency with \cite{gruppioni2013lf} and other SFRD estimates we use the total (AGN+galaxy) IR emission, although we stress the AGN is strongly sub-dominant, and a more quantitative analysis is postponed to a future stand-alone work. For these reasons, we do not remove the AGN contribution to the total $L_{\rm IR}$.
\par
\begin{figure*}[]
\centering
{\includegraphics[width=1.\textwidth]{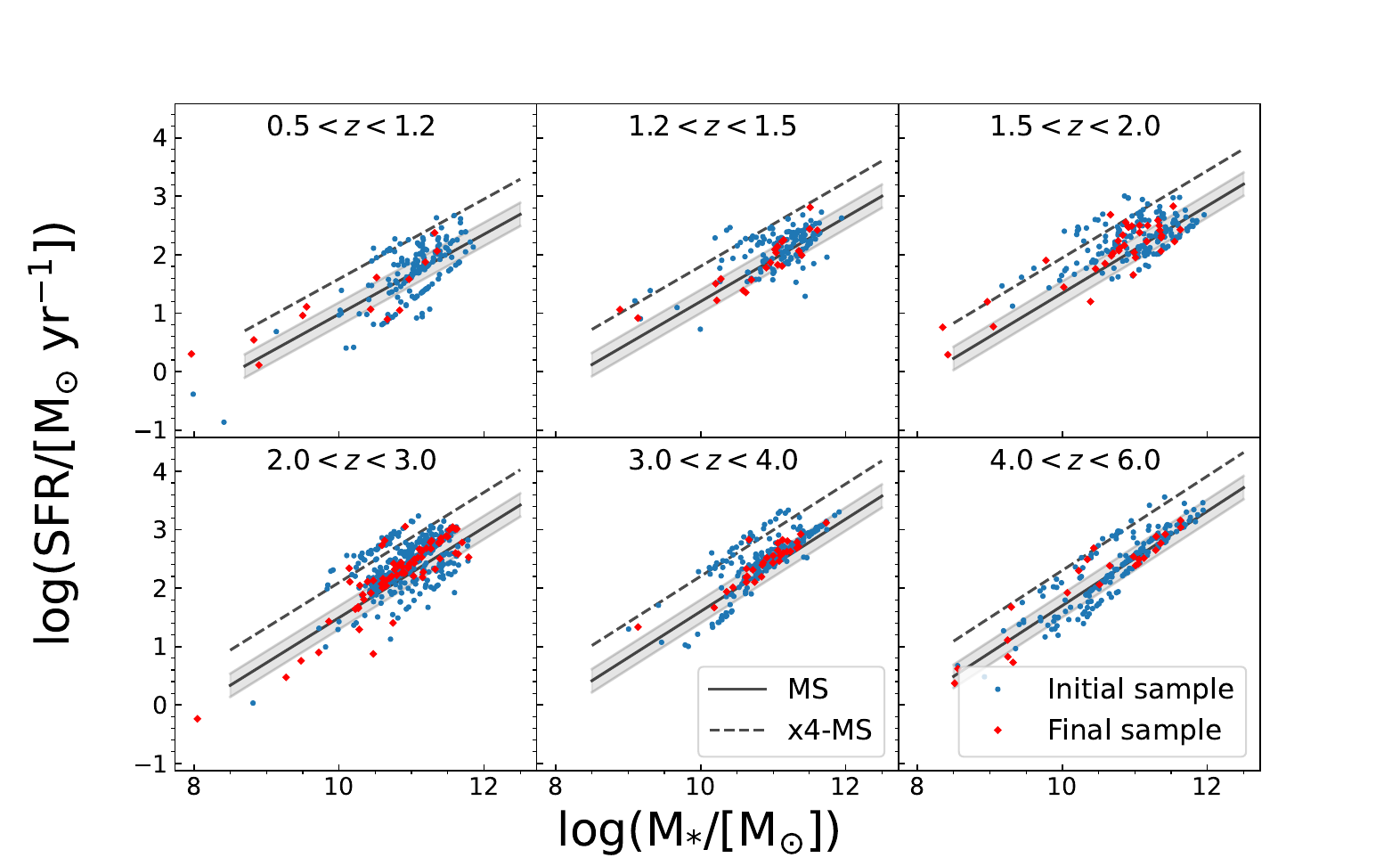}}
\caption{\small{Main sequence of galaxies computed in six different redshift bins: 0.5-1.2, 1.2-1.5, 1.5-2.0, 2.0-3.0, 3.0-4.0, 4.0-6.0. Data of the full sample are reported as blue circles, while the red diamonds represent the final sample (\ref{subsec:areal}) used in our analysis. The black solid line and shaded area indicate the main sequence \citep{speagle2014MS} (computed at the mean values of the redshift bins) and $1\sigma$ dispersion. Finally, the black dashed lines represent the 4x MS, indicative of the starburst regime.}}
\label{fig:sfr_z}
\end{figure*}

 Finally, we derived the dust obscured SFR for each galaxy using the \cite{kennicutt1998sfr} relation, which assumes the SFR to be proportional to the IR luminosity:
\begin{equation}
    \rm SFR \it (\rm M_{\odot} \rm yr^{-1}) \simeq 1.09 \times 10^{-10} L_{\rm IR}(\rm L_{\odot})
\end{equation}
We show the SFR against stellar mass distribution (main sequence, MS) in Figure \ref{fig:sfr_z}. Our sample is characterized by highly star forming galaxies, with SFRs up to $\sim 10^3$ $M_{\odot}$ yr$^{-1}$, typical of starbursting galaxies at the considered redshift. In each panel, as a black solid line, we plot the main sequence from \cite{speagle2014MS} computed at the mean redshift of the bin and an upper line (in black dashed) corresponding to 4 times the MS, which is indicative of the starbursting regime \citep{rodighiero2011MS}. Starburst galaxies tend to cluster on a “sequence” above and parallel to the MS, as also noticed in other works \citep{caputi2017ms,bisigello2018ms}. As our SFR is derived  directly from the IR luminosity, this bimodality can not be simply explained with the parametrization of the SFH or the presence of dust, but the exact origin of this effect needs further investigation with additional data. However, in our case, PIs selection can affect the MS distribution. In particular, we show in Figure \ref{fig:sfr_z} the sources distribution on the MS plane before (1411 sources) and after (189 sources) removing the targeted sources (i.e., the target of the individual pointing) in blue and red. As it can be noticed, the red points occupy mostly the region on the MS. 

%\subsubsection{Unfitted sources}\label{subsec:6outliers}
Finally, for 6 out of 189 sources, we were unable to obtain a reasonable fit. The common characteristic of these sources is that they have an SED characterized by optical and ALMA photometry that does not appear to belong to the same object and can only be fitted by assuming extreme, unrealistic templates of dust emission. For this reason, and by inspecting the ALMA and optical cutouts, we have decided to treat these sources statistically in the derivation of the LF. The details of the procedure, as well as some example, are shown in section \ref{subsec:method}.

%%%%%%%%%%%%%%%%%%%%%%%%%%%%%%%%
%
%Making the a3cosmos survey blind
%
%%%%%%%%%%%%%%%%%%%%%%%%%%%%%%%%

\section{Turning the \ad sample into a blind survey}\label{sec:blind} 

\begin{figure}[]
\centering
{\includegraphics[width=.5\textwidth]{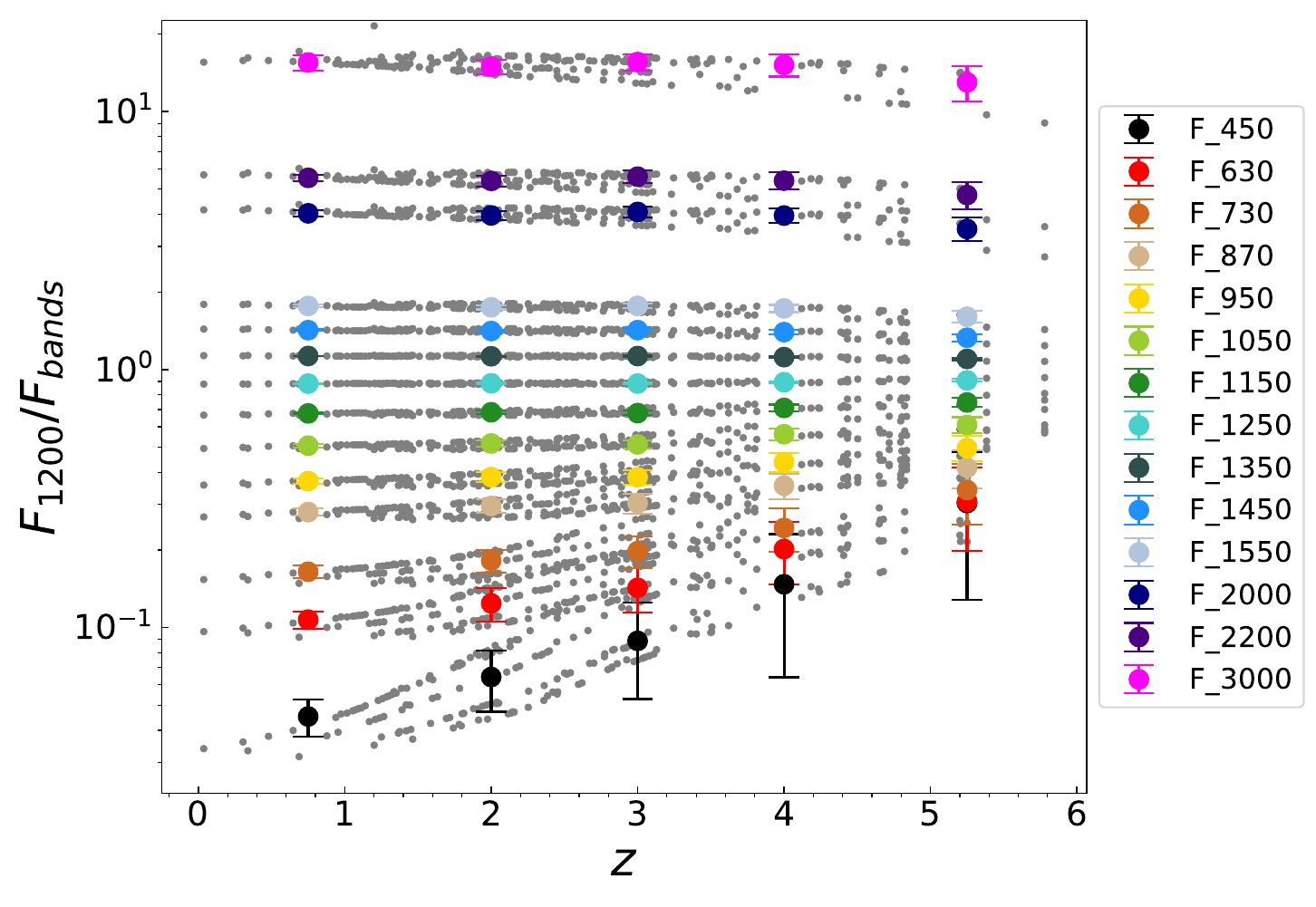}}
\caption{\small{Ratio between the flux at 1200 $\mu$m and other bands against redshift. Grey points on the background indicate the values for individual sources in each band, while colored points refer to the median values in each redshift bin and for each ALMA band, as described in Section \ref{subsec:conversion}.}}
\label{fig:f_r_z}
\end{figure}

\begin{figure}[]
\centering
{\includegraphics[width=.47\textwidth]{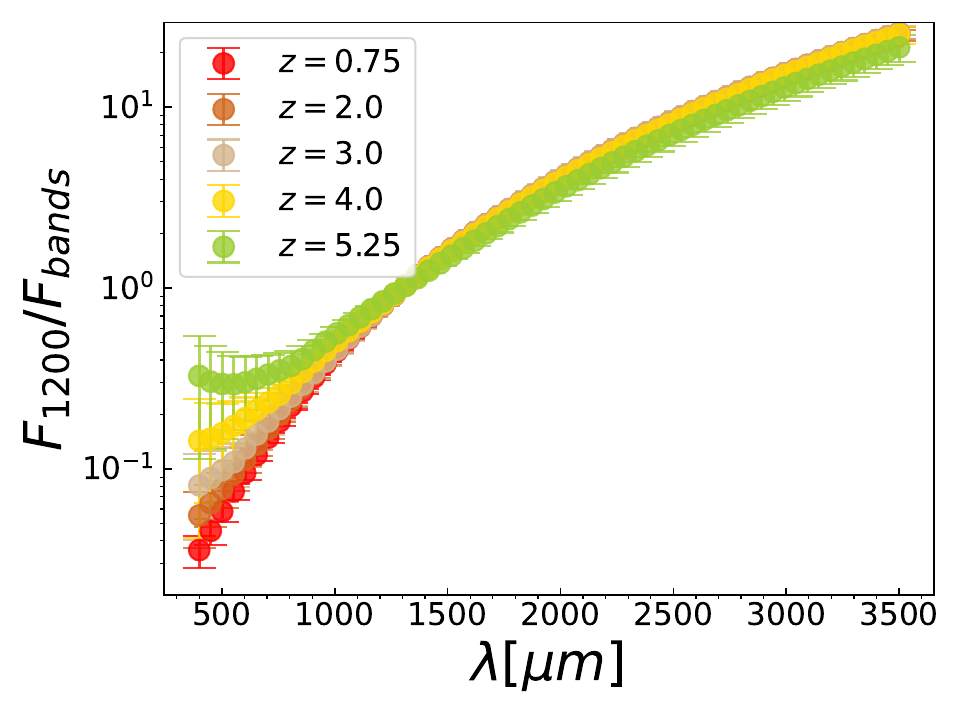}}
\caption{\small{Ratio between the flux at 1200$\mu$m and in other bands against observed wavelength. Each point is the median value computed as described in Section \ref{subsec:conversion}. Red, brown, grey and yellow points are at $z<4.5$, while the green ones represent the last redshift bin at $z>4.5$.}}
\label{fig:f_r_l}
\end{figure}

The \a3 survey is based on observations with different sensitivities (i.e., limiting fluxes), resolutions and ALMA observed-frame wavelenghts. Resulting in a survey of observations with unique selection functions. Being a collection of several observations, almost each pointing is centered on a targeted source of interest for that observation. In this perspective, the use of the \a3 survey for statistical purposes (e.g., LF, SFRD) needs a dedicated analysis for turning it into a blind survey. 
\par In this section, we discuss the method used to turn a generic (F)IR-mm ensemble of pointed observations into a blind survey (see also \textcolor{blue}{Adscheid et al.} in prep). 

\subsection{Blind surveys}

To derive statistical properties of the sample through the corresponding areal coverage, a blind survey is needed. In order to do this, the next step should be followed and features fulfilled:
\begin{itemize}
    \item In determining the limiting flux of A$^3$COSMOS, we scale the root mean square (RMS) of each individual pointing to the corresponding RMS at our reference wavelength of 1.2 mm (Section \ref{subsec:conversion});
    \item We determine the total area spanned by our survey, accounting for primary beam attenuation and overlapping pointings (Section \ref{avni});
    \item Finally, any possible bias due to the presence of a target has be taken into account (Section \ref{subsec:areal}).
\end{itemize}

%In particular, at first, all the observations that compose the survey should be homogenized in their observing wavelength, i.e., the root mean square (RMS) of all the pointings should be corrected to take into account their $\lambda_{\rm obs}$. In section \ref{subsec:conversion} we describe the procedure adopted to compute this correction factor. Secondly, we need to deal with the case in which two or more pointings are overlapping together. This is done by applying the method of \cite{A&B1980sample} and is described in Section \ref{avni}. Finally, any possible bias due to the presence of a target should be taken into account, this is done in Section \ref{subsec:areal}.

\subsection{Root mean square conversion between different $\lambda_{obs}$}\label{subsec:conversion}
The first step needed to achieve an unbiased survey from the \a3 sample is the conversion of all observing wavelengths to a reference one. In our case, we choose the $\lambda_{\rm ref} = $ 1200 $\mu$m wavelength in the observed frame, which falls in ALMA Band 6, being the most populated. This way, we can rescale all the fluxes at each observing wavelength ($\lambda_{\rm obs}$) to a reference one ($\lambda_{\rm ref}$), using the fluxes at the observed SEDs, in order to infer the observed ratio between $\lambda_{\rm ref}$ and $\lambda_{\rm obs}$ for each source. In particular we expect a decrease of the flux ratios when going to higher $z$ and shorter wavelengths, approaching the rest-frame peak of dust emission in the SED. 
\par As reported in Figure \ref{fig:f_r_z}, we divided the sample into five redshift bins (z<1.5, 1.5-2.5, 2.5-3.5, 3.5-4.5, 4.5-6). For visualization purposes, we plot only the most populated bands as grey points. The colored points indicate the median value for a certain $\lambda_{obs}$ for a specific redshift bin. It can be seen that, for shorter $\lambda_{obs}$ (close to $\lambda_{\rm ref}$) the ratios are unsurprisingly low and almost redshift independent, while at longer $\lambda_{\rm obs}$, the ratios become larger and more redshift-dependent, especially in the highest $z$-bin. 
\par Figure \ref{fig:f_r_l} shows the differences between the ratios in the different $z$-bins. We note that for $z<4.5$, the correction curves are very similar to each other. For this reason we used only one mean curve (up to $z=4.5$) and a different one for $z>4.5$ in our analysis. 

\subsection{Radial limiting flux} \label{avni}
We used the conversion curves obtained as explained in the previous section to convert the root mean square (RMS) in each pointing as if it was observed at 1200 $\mu$m. The pointings are now characterized by a uniform (in wavelength) sensitivity across their field of view (FoV). However, observing with the ALMA interferometer leads to a primary beam in which the sensitivity of the observation varies radially from the center to the outer regions. 
\par For this reason, we divided the sky region covered by our pointings into 1" pixels and flagged those inside a pointing with a flux value corresponding to the sensitivity of that pointing (i.e., primary beam correction higher than 0.2).
%The pixels found inside a pointing are flagged with a flux value corresponding to the sensitivity of that pointing, while we associate an arbitrarily large flux value (10$^6$ $\mu$Jy) to pixels lying outside the pointings (i.e., primary beam correction lower than 0.2). 
Then, we can parametrize the primary beam correction ($pbcor$ hereafter), as a Gaussian function, peaking at the center of the circle, with a value of 1, decreasing to 0 towards the outer regions:
\begin{equation}
    pbcor = e^{-\frac{d^2}{2\sigma^2}},
\end{equation}
with $d$ being the distance of a pixel from the center of the pointing and $\sigma$ being the $FWHM/2.35$. 
\par By dividing pixel by pixel the RMS by the $pbcor$ corresponding to that pixel distance, we obtained a corrected RMS, which is then converted to the limiting flux by multiplying it by 4.35, which is the sensitivity cut of the \a3 prior catalog.
After this procedure we end up with a pixel map of limiting fluxes inside the pointings. However, for the pointings that are overlapping, the selection of an exclusive area is not straightforward. Indeed, the common area between two overlapping pointings has to be counted only once, in which a limiting flux from one of the two pointings needs to be assigned. 
\par
\begin{figure} 
\centering
\begin{minipage}[b]{.45\textwidth}
\centering
\includegraphics[width=1\textwidth]{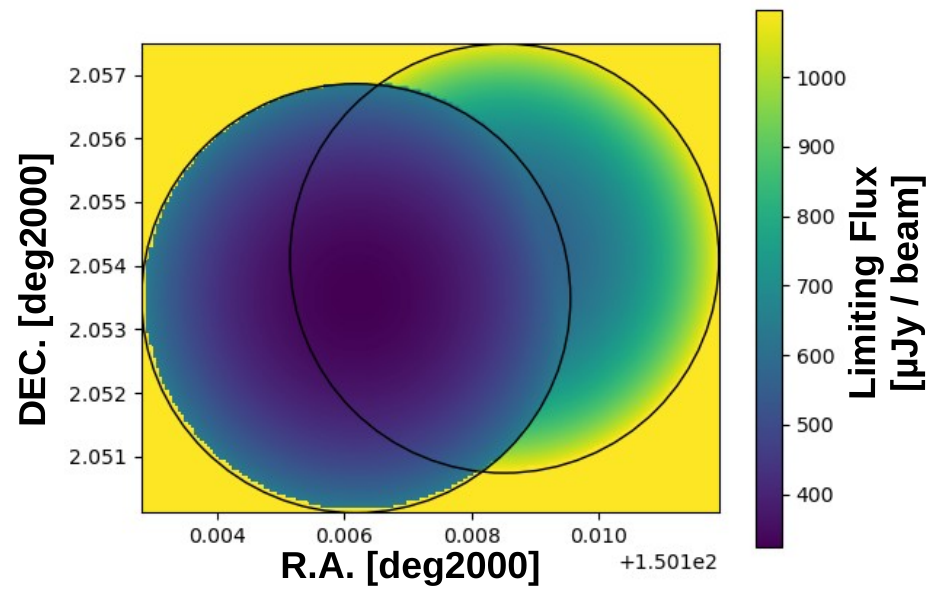}
\end{minipage}
\caption{Two overlapping pointings in band 6. The black circle represents the adopted primary beam size (out to a $pbcor$ of 0.2), being color-coded by the inferred RMS. Lower RMS values are represented by darker colors and higher values by lighter ones. Outside our primary beam boundary, we have set values to the arbitrary high value of $10^6$ $\mu$Jy. In these type of cases, we adopted the area from the deepest pointing, in the overlapping region.}
\label{fig:overlapping}
\end{figure}

 This issue can be dealt with by following the \cite{A&B1980sample} method which coherently combines multiple samples at different depths. %In fact, for each source in the sample, a corresponding comoving volume of the source is needed to study statistical properties like luminosity function or luminosity density. 
%However, if a source is present in more than one pointing, it has to be counted as a single source, but with different possibly available volumes, each corresponding to a certain limiting flux. %The \cite{A&B1980sample} method allows us to assign to each overlapping pixel the correct limiting flux. %In particular, since in our survey the sensitivity would have allowed that source to be observed up to the maximum volume available for the source, in a given pointing, then, this information must be used to assign the correct limiting flux to an overlapping region.
As shown in Figure \ref{fig:overlapping}, the limiting flux in the common area covered by two or more pointings is the one corresponding to the most sensitive observation at the reference frequency (i.e., the lowest limiting flux among the pointings). %We show in Figure \ref{fig:overlapping} and example of two overlapping pointings with different limiting fluxes. Firstly, it can be noticed that the fluxes are varying radially from the center to the outer regions, meaning that the observations are deeper in the central part and shallower in the outer limits. Also, the effect of applying the Avni \& Bachall approach can be observed. Indeed, the overlapping region is characterized by the limiting flux of the deeper pointing (the one on the left) and then, a source observed in both pointings, will be characterized by a volume corresponding to that limiting flux.

\subsection{Areal coverage} \label{subsec:areal}
We proceed with deriving the cumulative areal coverage of our survey, by combining the effective area accessible by all pointings, at each limiting flux. This is done by counting the number of pixels with a certain limiting flux and deriving their cumulative distribution function (CDF). In this way, it is possible to associate an area to a limiting flux ($S_{\rm LIM}$), which is the portion of area having a flux lower than $S_{\rm LIM}$. 
%\par Since we have two different conversion curves for the RMS, we have to compute the areal coverage curves derivation for two different pointings map (the difference is only on the pixel-by-pixel limiting flux) and obtain the areal coverage curves for $z<4.5$ and $z>4.5$.
%The areal coverage are reported in Figure \ref{fig:areal_coverage}. \textbf{Aggiornare e descrivere la figura}.
%\par Finally, we need to exclude the targeted sources \textbf{cosa posso scrivere qua, anche in merito agli ultimi aggiornamenti di Sylvia con in number counts etc.. ?}
\par In order to minimize potential biases due to pre-targeted ALMA sources contaminating the selection function, we disregard central targets as follows. We removed pointings in which a target source is not present in the central $1"$ (2060/3215). In this way, we took into account the possible positional offset of SMG-galaxy follow up with ALMA, which is of about $5"$ and, therefore, could be wrongly addressed as a serendipitous detection. By removing pointings without a target, we are sure to not bias our sample with possible off-center targeted galaxies. The $1"$ masking has been chosen through simulation varying the mask radius, increasing its size, and
comparing the retrieved masked number counts with an input distribution (\textcolor{blue}{Adscheid et al., prep.}). It has been shown that by increasing the central mask radius, no benefit is observed in the convergence of the retrieved number counts.
In addition, we also removed 104/1620 sources that are potentially clustered (i.e. having a similar redshift to the target) following the criterion from \cite{weaver2022cosmos2020}, for which a source can be considered at the same redshift of another one if the difference in module between redshifts ($z_s$ and $z_t$) is smaller than $0.04(1 + z_t)$. Finally, we removed the target of each pointing.
\par
\begin{figure}[]
\centering
{\includegraphics[width=.45\textwidth]{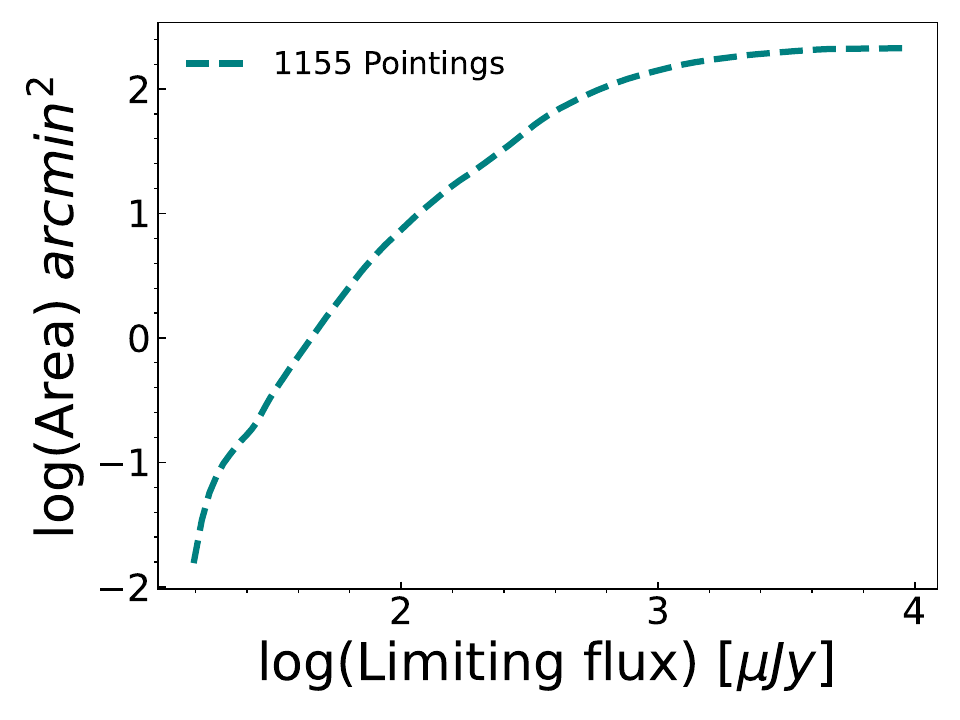}}
\caption{\small{Total areal coverage of the 1155 pointings after the cut for the lack of target detection within 1 arcsec.\\}}
\label{fig:areal_coverage}
\end{figure}

 The final sample, obtained by considering both the target and redshift cuts, consists of 189 sources out of our initial sample of 1620 sources. We show in Figure \ref{fig:areal_coverage} the areal coverage for the remaining 1155 pointings, i.e, after removing pointings without a central target detection.

%%%%%%%%%%%%%%%%%%%%%%%%%%%%%%%%
%
%ALTRO?
%
%%%%%%%%%%%%%%%%%%%%%%%%%%%%%%%%

\section{\a3 luminosity function}\label{sec:lf}
Deriving the areal coverage of our survey allows us to properly compute the luminosity function with the $1/V_{\rm MAX}$ method \citep[][]{Schmidt1970vmax}. In the following sections we describe the method applied to derive the total infrared luminosity function and compare it with previous works.
\subsection{The method}\label{subsec:method}
We derived the LF using the \cite{Schmidt1970vmax} method, that, based on the data, allows us to derive the LF without making any assumption relative to the LF shape. As already mentioned in section \ref{sec:blind}, the \a3 survey is composed of $\sim 1155$ individual pointings (after the cut), that can be considered as independent fields. For this reason, by following the \cite{A&B1980sample} method, we are able to derive the effective areal coverage at each $S_{\rm LIM}$, across all pointings. The relation between area and limiting fluxes (see Figure \ref{fig:areal_coverage}) is then used to associate an accessible area above a certain flux for each source. % curve that takes into account each pointing. The area-limiting flux curve is then used to associate a covered area to a certain flux in each position inside the pointings. 
\par To compute the LF, we divided our sample in eight redshift bins, from $z\sim 0.5$ to $z\sim 6$ and into luminosity bins of 0.5 dex width, from log($L_{\rm IR}/\rm L_{\odot})=10$ to log($L_{\rm IR}/\rm L_{\odot})=14$. For each source, in a $z$-$L_{\rm IR}$ bin, we measure the contribution to the LF in that bin by applying a redshift step of $dz=0.02$ and $K$-correcting its SED from the lower to the upper boundary of the corresponding redshift bin, each time computing the observed flux at 1200 $\mu$m. This flux is used to infer the corresponding areal coverage at each $dz$ by interpolating the previously derived areal coverage curve (Figure \ref{fig:areal_coverage}). We finally combine the effective area obtained in this way with the element of volume at each redshift step and obtain a comoving volume over which a given source is accessible:
\begin{equation}
    V_{\rm MAX} = V_{\rm zmax} - V_{\rm zmin}, 
\end{equation}
where $V_{\rm zmax}$ and $V_{\rm zmin}$ are the sum of the sub-volume in each $dz$ shell up to the upper and lower limit of the bin, respectively. In particular $V_{\rm zmax}$ can either be the volume at the upper bound of each $dz$ bin, or the maximum volume reachable by considering the $S/N$ limit of the survey (i.e., corresponding to the $z$ at which the area would be 0).
%\par The $V_{\rm MAX}$ is then corrected by a factor depending on the number of missing sources in each redshift bin (i.e., sources without a well fitted SED). %Also, sources without a redshift have been assumed to be distributed in the bins at $z>3$, assuming a similar distribution in number as to the other sources:
%\begin{equation}
    %V_{\rm MAX,c} = V_{\rm MAX} / C_{\rm obj},~~ C_{\rm obj} = \frac{N_{\rm obj,bin}}{N_{\rm obj}}, 
%\end{equation}
%with $N_{\rm obj,bin}$ being the total number of sources belonging to that $z$-bin and $N_{\rm obj}$ the actual number of sources we were able to use, in the same bin, i.e., sources with a photometric or spectroscopic redshift in this redshift bin and with a good \texttt{CIGALE} fit. 
Finally, we correct the $V_{\rm MAX}$ by taking into account completeness and spuriousness corrections, derived by \cite{liu2019a31}, and we obtain the $\Phi(L,z)$ by summing each $1/V_{\rm MAX}$ in a certain luminosity-redshift bin.

\subsection{Statistical treatment for potential HST-dark sources}
As described in Section \ref{subsec:sedres}, we were unable to obtain a satisfactory fit for the SED of 6 out of 189 sources. Additionally, we identified these 6 sources as potential HST-dark objects. Consequently, due to having only ALMA flux information (the photo-$z$ is associated with optical photometry), these objects were treated statistically in the LF analysis. Specifically, we assumed that these sources exist at a redshift greater than 3 and followed this procedure:
\begin{itemize}
 \item firstly, we computed the cumulative redshift distribution for the rest of the sample;
 \item next, for each $N$-th ($N=100$) random extraction during the LF calculation, we assigned a redshift to each of the sources by drawing from the cumulative distribution function (CDF) as a random sampler;
 \item once a redshift was drawn, we used the median SED of the sample to “fit” the ALMA flux and, consequently, obtain an infrared luminosity to be incorporated into the LF calculation.
\end{itemize}

\begin{figure}[]
\centering
{\includegraphics[width=.5\textwidth]{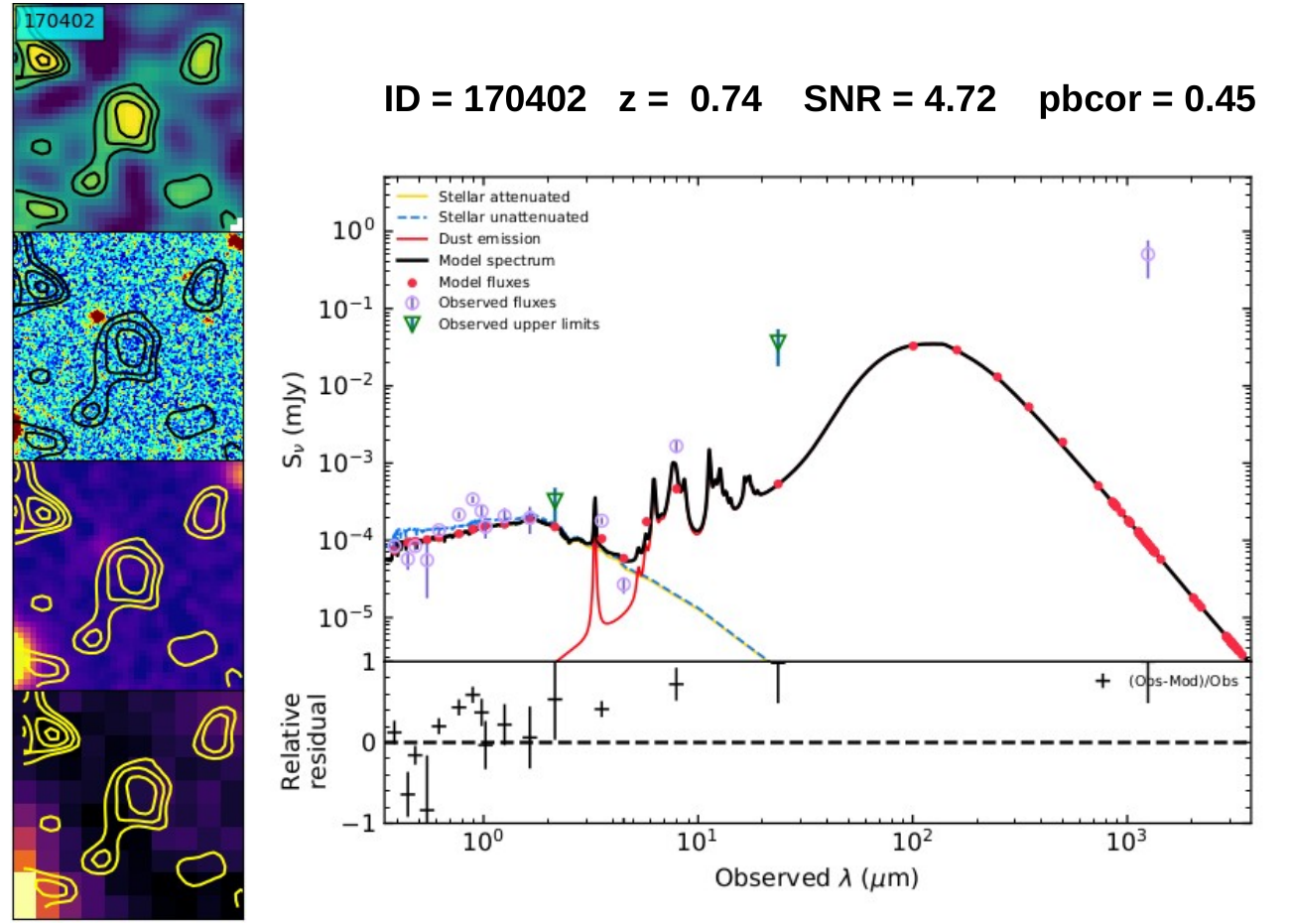}}
\caption{\small{Example of a source with optical identification (photo-$z$ = 0.74) likely distinct from the ALMA galaxy. Left panel (from top to bottom): ALMA, acs-I, UV-j, and IRAC1 images are displayed with ALMA contours overlaid. It can be inferred that the optical/NIR object near the ALMA position is not centered in the ALMA galaxy. On the right: SED of the same object, showing the impossibility of fitting the optical and ALMA mm flux simultaneously.}}
\label{fig:hstdark}
\end{figure}

\subsection{IR LF}\label{subsec:totLF}

\begin{figure*}[]
\centering
{\includegraphics[width=1.\textwidth]{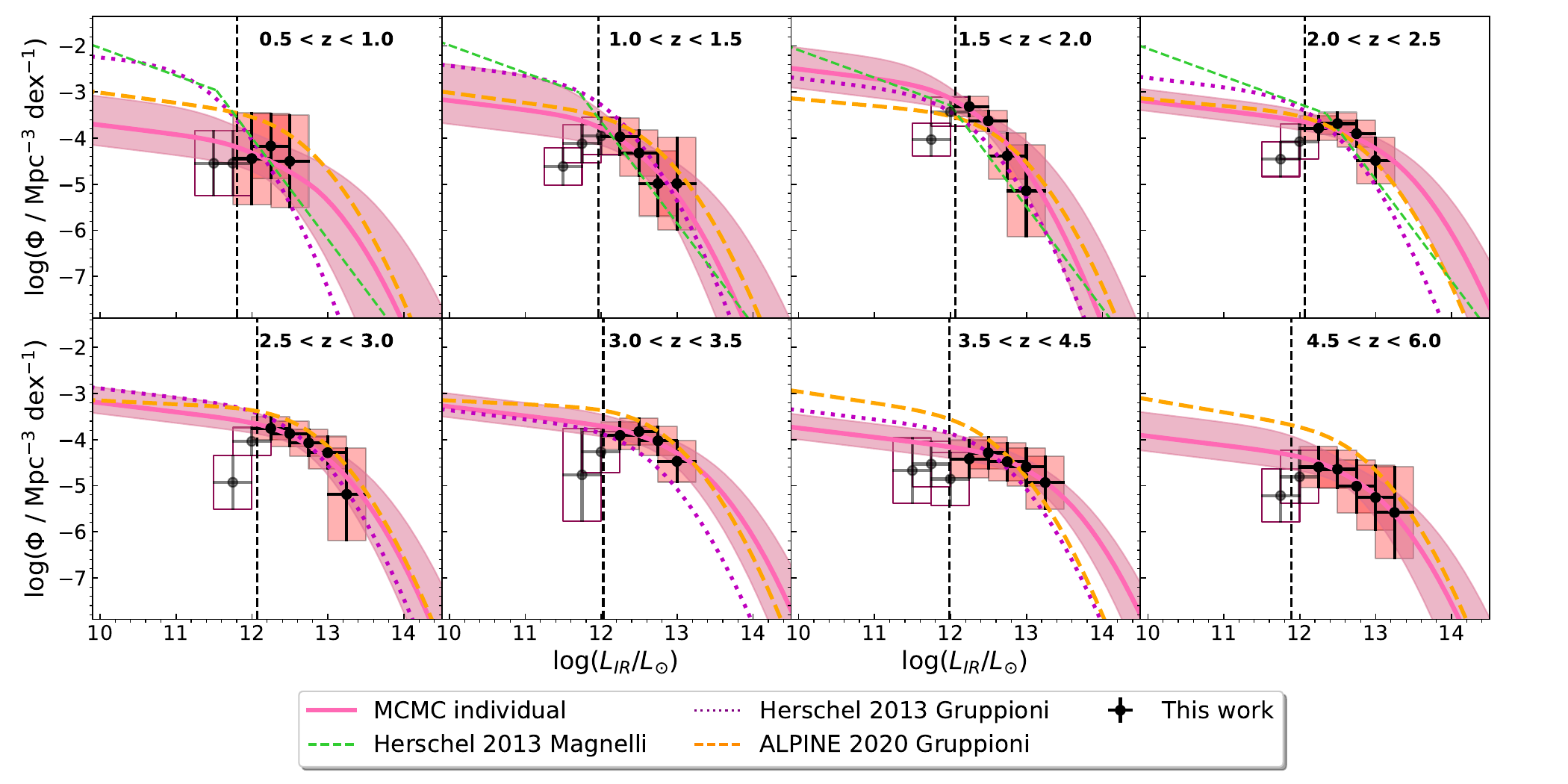}}
\caption{\small{A$^3$COSMOS luminosity function (black points and red boxes). The individual redshift bin MCMC best-fit is plotted as pink solid lines and shaded error bands. The redshift ranges are reported in the upper-right corner of each subplot, while the luminosity bins are centered at each 0.25 dex, with a width of 0.5 dex (overlapping bins). The black vertical dashed lines represent the completeness limit of the L$_{IR}$. The orange dashed, purple dotted and green dashed lines are the best fit LF obtained by \cite{gruppioni2013lf,magnelli2013ir,gruppioni2020alpine}.}}
\label{fig:LF_models_a3_only}
\end{figure*}

\begin{figure*}[]
\centering
{\includegraphics[width=1.\textwidth]{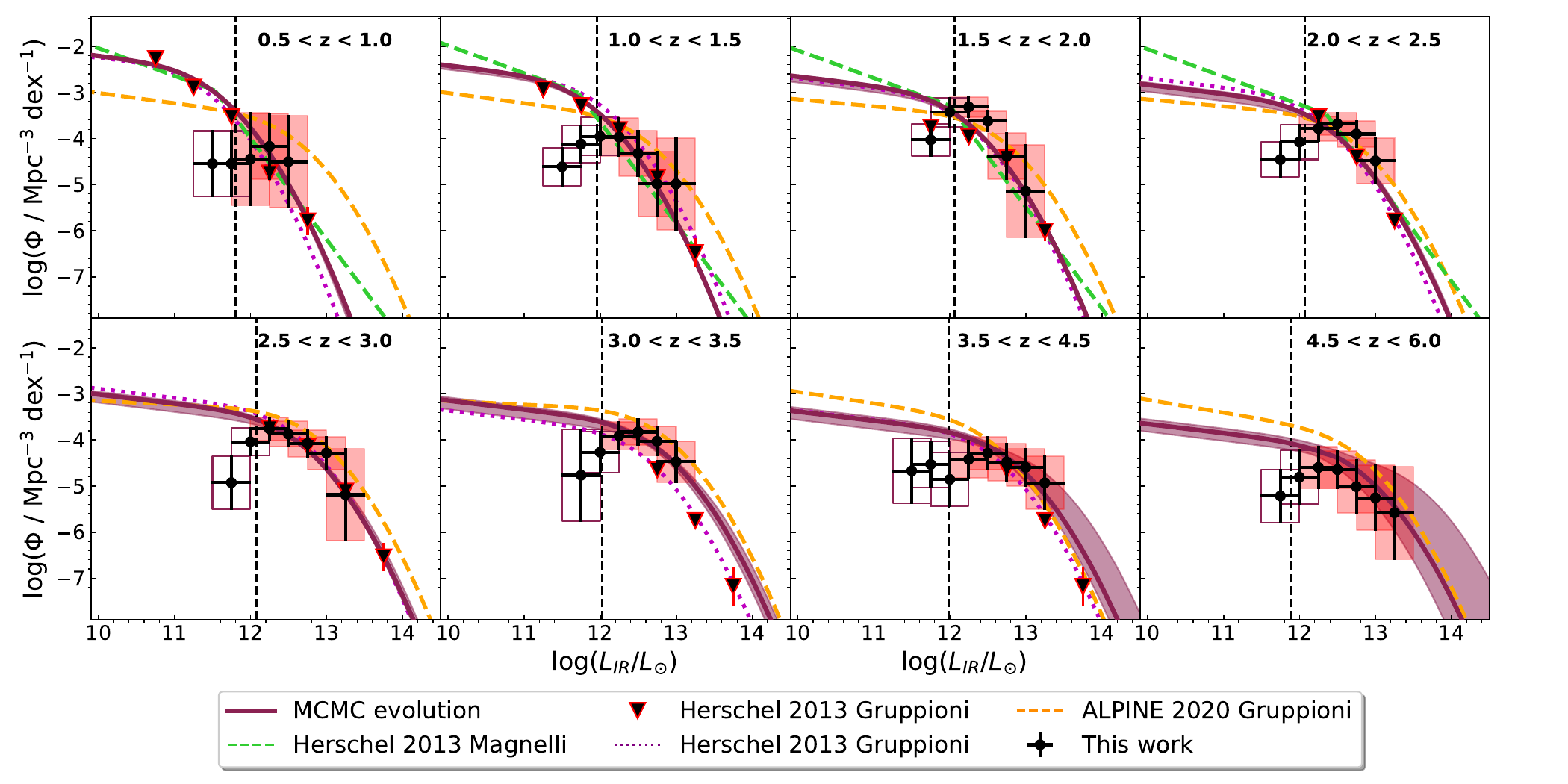}}
\caption{\small{A$^3$COSMOS (red boxes and black points) + \textit{Herschel} (black triangles) luminosity function. The dark-red lines and shaded areas are the best-fit obtained by using all the LF point from different $z$-bins together. The redshift ranges are reported in the upper-right corner of each subplot, while the luminosity bins are centered at each 0.25 dex, with a width of 0.5 dex (overlapping bins). The black vertical dashed lines represent the completeness limit of the $L_{\rm IR}$. The orange dashed, purple dotted and green dashed lines are the best fit LF obtained by \cite{gruppioni2013lf,magnelli2013ir,gruppioni2020alpine}.}}
\label{fig:LF_models}
\end{figure*}

We obtained the IR LF by using the total (i.e., including all the SED components) $L_{\rm IR}$ computed in the $8-1000$ $\mu$m range. %The SF-related L$_{IR}$ will be used later (Section \ref{sec:SF-only}) to assess AGN contamination to the total IR luminosity and to purify the SFRD (Figure \ref{fig:sfrd_comparison}).
The $z$-bins (0.5-1.0; 1.0-1.5; 1.5-2.0; 2.0-2.5; 2.5-3.0; 3.0-3.5; 3.5-4.5; 4.5-6.0) are nearly equally populated, apart from the first bin which contains slightly less sources than the others. 
The LF is shown in Figure \ref{fig:LF_models_a3_only} as black points and red boxes representing the Poissonian uncertainties and the values are reported in Table \ref{tab:lf}. In each bin we overplot the completeness limit as a vertical black dashed line. This threshold is computed by re-scaling all the observed 1200$\mu$m fluxes of each SED to the same limiting flux, and then taking the $L_{\rm IR}$ of SED which gives the highest $L_{\rm IR}$ value at that limit. This latter value represents the value of $L_{\rm IR}$ below which our sample is not 100$\%$ complete. The points below the completeness limit are reported as empty boxes. We also report existing estimates obtained from other studies. In particular, we compared our results with best fit from previous IR works, either from ALMA \cite[ALPINE,][]{gruppioni2020alpine} or Herschel \citep[PEP+GOODS-H and PEP+HerMES,][]{magnelli2013ir,gruppioni2013lf}, when available for redshifts that are similar to ours.
\par As it can be seen from Figure \ref{fig:LF_models_a3_only}, we are mostly sampling the bright-end of the luminosity function, with our complete points being above log($L_{\rm IR}/ \rm L_{\odot}$) $\simeq 12$ at each redshift bin. This is mainly a consequence of the fact that, at lower redshift, ALMA is sampling down the Rayleigh-Jeans tail and therefore even if deep pointing are available, they are inefficient compare to Herschel which sample the peak of the IR SED. The peak of dust emission can be better probed by ALMA (band 6 and 7 mainly) at high redshifts ($z > 3$). Despite this, our data points are in good agreement, within the errors, with the black triangles from Herschel (see Figure \ref{fig:LF_models}), whose $z<3$ LF estimates are characterized by much higher statistics (smaller errorbars) and better sensitivity, being able to probe the IR LF down to the faint-end and the knee region (log($L_{\rm IR}$/$ \rm L_{\odot}$) $\simeq 11$). From this consistency with the LF from \cite{gruppioni2013lf} we can assert that, despite the poorer statistics in terms of the number of sources, the method used to turn the \a3 into a blind survey is valide and accurate. At $z>3-3.5$, where Herschel is probing further down the peak of the dust emission while ALMA start probing the peak, our estimates show a systematically higher normalization. Indeed, as stated by \cite{gruppioni2013lf}, in the $3 < z < 4.2$ Herschel redshift bin, most of the sources have a photometric redshift and the PEP selection may be missing a fraction of high redshift galaxies, making this estimate a lower limit.
%\par We also compared our results with those derived from CO LFs \citep[ASPECS, COLDz,][]{decarli2019co, riechers2019co}, converted to L$_{IR}$ (grey and green boxes). The conversion factors adopted are those from \cite{carilli2013co} (log$L_{\rm IR}$=1.37log$L'_{\rm CO(1-0)}$-1.74) and \cite{decarli2016co} (log$L'_{\rm CO(1-0)}$=log$L'_{\rm CO(2-1)}$-log(0.76)). We observe that our LF data points are not consistent with the CO LF at $1<z<1.5$ and $3.5<z<4.5$ in the highest luminosities. At redshifts $2.0<z<2.5$ and $2.5<z<3.0$ our results are consistent within errors in the luminosity range covered wit them.

\begin{table*}[]
\centering

\renewcommand{\arraystretch}{1.5}
\caption{IR luminosity function as inferred using the \a3 database.}
\begin{threeparttable}
\begin{tabular}{ccccccccc}
\hline
\hline
log($L_{\rm IR}/L_{\odot})$ & \multicolumn{4}{c}{log($\Phi/\rm Mpc^{-3} \rm dex^{-1}$)}                                                                                     \\ \hline
                        & $0.5<z  \le 1.0$                           & $1.0<z \le 1.5$                            & $1.5<z \le 2.0$                            & $2.0< z \le 2.5$                            \\ \cline{2-5} 
11.25-11.75             & (\textit{-4.54} $\pm$ \textit{0.71 [3]})   & (\textit{-4.61} $\pm$ \textit{0.41 [8]})   &                                            &                                                         \\
\textbf{11.50-12.00}    & (\textbf{-4.54} $\pm$ \textbf{0.71 [2]})   & \textbf{(-4.12} $\pm$ \textbf{0.41 [9])}   & \textbf{(-4.03} $\pm$ \textbf{0.35 [9])}   & \textbf{(-4.46} $\pm$ \textbf{0.38 [7])}                \\
11.75-12.25             & \textit{-4.44} $\pm$ \textit{1.00 [1]}     & (\textit{-3.95} $\pm$ \textit{0.41 [9]})   & (\textit{-3.43} $\pm$ \textit{0.32 [12]})  & (\textit{-4.08} $\pm$ \textit{0.38 [7]})                \\
\textbf{12.00-12.50}    & \textbf{-4.17} $\pm$ \textbf{0.71 [2]}     & \textbf{-3.97} $\pm$ \textbf{0.41 [8]}     & \textbf{-3.31} $\pm$ \textbf{0.22 [21]}     & \textbf{-3.79} $\pm$ \textbf{0.27 [14]}                 \\
12.25-12.75             & \textit{-4.50} $\pm$ \textit{1.00 [1]}     & \textit{-4.32} $\pm$ \textit{0.50 [5]}     & \textit{-3.62} $\pm$ \textit{0.22 [20]}    & \textit{-3.68} $\pm$ \textit{0.24 [17]}                 \\
\textbf{12.50-13.00}    &                                            & \textbf{-4.99} $\pm$ \textbf{0.71 [2]}     & \textbf{-4.38} $\pm$ \textbf{0.5 [4]}      & \textbf{-3.90} $\pm$ \textbf{0.29 [12]}                 \\
12.75-13.25             &                                            & \textit{-4.99} $\pm$ \textit{1.00 [1]}     & \textit{-5.14} $\pm$ \textit{1.00 [1]}     & \textit{-4.48} $\pm$ \textit{0.50 [4]}                  \\ \hline
\multicolumn{1}{l}{}    & $2.5<z \le 3.0$                            & $3.0<z \le 3.5$                            & $3.5<z \le 4.5$                            & $4.5\textless{}z \le 6.0$ \\ \cline{2-5} 
11.25-11.75             &                                            &                                            & (\textit{-4.67} $\pm$ \textit{0.71 [2]})   & \textit{}                                               \\
\textbf{11.50-12.00}    &  \textbf{(-4.93} $\pm$ \textbf{0.58 [3])}  & \textbf{(-4.76} $\pm$ \textbf{1.00 [1])}   & \textbf{(-4.53} $\pm$ \textbf{0.50 [3])}   & \textbf{(-5.21} $\pm$ \textbf{0.58 [2])}               \\
11.75-12.25             & (\textit{-4.04} $\pm$ \textit{0.30 [11]})  & (\textit{-4.27} $\pm$ \textit{0.45 [3]})   & (\textit{-4.85} $\pm$ \textit{0.58 [2]})   & (\textit{-4.81} $\pm$ \textit{0.58 [3]})                  \\
\textbf{12.00-12.50}    & \textbf{-3.76} $\pm$ \textbf{0.25 [16]}    & \textbf{-3.91} $\pm$ \textbf{0.30 [9]}     & \textbf{-4.42} $\pm$ \textbf{0.41 [6]}     & \textbf{-4.60} $\pm$ \textbf{0.45 [5]}                  \\
12.25-12.75             & \textit{-3.87} $\pm$ \textit{0.28 [13]}    & \textit{-3.83} $\pm$ \textit{0.29 [13]}    & \textit{-4.29} $\pm$ \textit{0.35 [8]}     & \textit{-4.64} $\pm$ \textit{0.41 [6]}                  \\
\textbf{12.50-13.00}    & \textbf{-4.07} $\pm$ \textbf{0.29 [12]}    & \textbf{-4.03} $\pm$ \textbf{0.32 [11]}    & \textbf{-4.48} $\pm$ \textbf{0.41 [6]}     & \textbf{-5.01} $\pm$ \textbf{0.58 [3]}                  \\
12.75-13.25             & \textit{-4.28} $\pm$ \textit{0.35 [8]}     & \textit{-4.47} $\pm$ \textit{0.45 [5]}     & \textit{-4.59} $\pm$ \textit{0.41 [7]}     & \textit{-5.25} $\pm$ \textit{0.71 [2]}                  \\
\textbf{13.00-13.50}    & \textbf{-5.19} $\pm$ \textbf{1.00 [1]}     &                                            & \textbf{-4.93} $\pm$ \textbf{0.58 [4]}     & \textbf{-5.58} $\pm$ \textbf{1.00 [1]}                  \\ \hline \\
\end{tabular}
\begin{tablenotes}
   \small{ \item[*]{In the first column the luminosity bins are reported, while columns 2-8 report the $\Phi$ values in each luminosity and redshift bins. Bold (or italic) values represent independent luminosity bins. Values in round brackets indicate luminosity bins which are below the completeness limit and numbers in square brackets are the number of sources in each L-$z$ bin.}}  
\end{tablenotes}
\end{threeparttable}
\label{tab:lf}

\end{table*}

\subsection{LF best-fit and evolution}\label{mcmc}
In order to trace the number density of galaxies at different redshifts and infrared luminosities, we quantify the LF using a Schechter function. To this purpose, we modeled our LF with a modified Schechter function \citep[][]{saunders1990modified}, described by four free parameters:
\begin{equation}
    \Phi(L)dlogL = \Phi^*\left(\frac{L}{L^*}\right)^{1-\alpha_S} exp\left[-\frac{1}{2\sigma_S^2}log_{10}^2 \left(1+\frac{L}{L^*}\right)\right]dlogL,
\end{equation}
where $\alpha_S$ and $\sigma_S$ are the slope of the faint end and the parameter shaping the bright end slope, respectively, and $L^*$ and $\Phi^*$ represent the luminosity and normalization at the knee, respectively. The modified Schechter function is similar to a power law for $L \ll L^*$ and behaves as a Gaussian for $L \gg L^*$.  
\par To find the $L^*$ and $\Phi^*$ that best reproduce our LF, we performed Monte Carlo Markov Chain (MCMC) analysis using the \texttt{PYTHON} package \texttt{emcee} \citep{foremanmackey2013emcee}. It uses a set of walkers to explore the parameter space simultaneously. We carried out the MCMC analysis using 50 walkers, with 10000 steps (draws) and discarding the first 1000 sampled draws of each walker (burnin). The likelihood has been built in the form:
\begin{equation}
    L = -\frac{1}{2} \sum \left(\frac{\Phi_{\rm Model} - \Phi}{\delta \Phi}\right)^2 .
\end{equation}
 
We ran the MCMC using flat prior distributions for the two free parameters and with $\alpha_S$ and $\sigma_S$ fixed to the values of \cite{gruppioni2013lf} (i.e., $\alpha_{S}$=1.2 and $\sigma_{S}$=0.5), with log($\Phi^*$) between $-5$ and $-2$ and log($L^*$) between 10 and 13. The prior distribution is then combined with the likelihood function to obtain the posterior distribution.
\par Firstly, we fitted the ALMA points alone (Figure \ref{fig:LF_models_a3_only}); in the lower redshift bins (i.e., $0.5<z<1.0$ and $1.0<z<1.5$), the individual fit shows a very low normalization with respect to the Herschel best-fit and large error bands, which do not allow us to constrain the fit parameters $L^*$ and $\Phi^*$ in an accurate way. Between redshift 1.5 and 3.0 the best-fit LF is in a good agreement with Herschel, except for a slightly higher, but consistent, normalization in the $1.5<z<2.0$ redshift bin. Finally, between $z=3$ and $z=3.5$, our best-fit is consistent with that of the ALPINE survey \citep[][]{lefevre2020alpine}, but has lower $\Phi^*$ at knee at $z>3.5$. 
\par Since the ALMA-only LF (red boxes) is not able to trace the lowest luminosities, we decided to take advantage of the great statistic and coverage in luminosity of the PEP+HerMES survey by Herschel \citep[][]{gruppioni2013lf} up to $z \sim 3$, with which our sample is consistent, while extending to higher redshift using our ALMA measurements. By means of this, it is possible to better characterize the faint-end and the knee of the LF at $z>3$. Indeed, by using Herschel's LF data at all available redshifts (in Figure \ref{fig:LF_models}, only those within the redshift bins from 0.5 to 6 are shown), the MCMC analysis we conducted manages to better characterize the parameters of the Schechter function and subsequently trace its evolution at higher $z$.
%At first, we performed the MCMC by fitting each redshift bin separately, in order to derive $L^*$ and $\Phi^*$ only from the data present in that bin. The best-fit at each redshift is plotted in Figure \ref{fig:LF_models} as a red solid line, with red shaded area being the 16th and 84th percentiles of the error distribution computed by computing 100 iterations of the IR LF and MCMC fit, and by using $L_{\rm IR}$ extracted from a Gaussian centered in the best-fit $L_{\rm IR}$ of each galaxy. At redshifts $z<3$, where we included the Herschel data to the fit, the result is very similar to the purple dotted line (note that the purple dotted line refers to slightly different central redshift values). At $z>3$, since we are not using Herschel, the red line refers to the \a3 only best-fit. The best fit parameters ($L^*$ and $\Phi^*$) are shown in Table \ref{tab:lf_mcmc}. 
\par Previous works have already pointed toward a redshift evolution of both typical density and luminosity of the IR galaxy population \citep[see e.g.,][]{caputi2007lfevol,bethermin2011lfevol,marsden2011lfevol,gruppioni2013lf}, characterized by an increase in luminosity and decrease in density with increasing redshift. In particular both \cite{caputi2007lfevol, bethermin2011lfevol} and \cite{gruppioni2013lf} found a break in the redshift evolution, resulting in a steepening of the density evolution a $z>1$ and a flattening for the luminosity at $z>2$. For these reasons, we performed a second MCMC fit using the multi-redshift information from each redshift bin,
assuming an exponential shape and two different $z_{break}$ values ($z_{\rho0}$ and $z_{l0}$), for the evolution of $\Phi(z)^*$ and $L(z)^*$, expressed as:

\begin{equation}
\begin{cases}
  \Phi^* = \Phi^*_{0} (1+z)^{k_{\rm \rho1}} ~~~~~~~~~~~~~~~~~~~~~~~~ z < z_{\rm \rho0}\\
  \Phi^* = \Phi^*_{0} (1+z)^{k_{\rm \rho2}} (1+z_{\rm \rho0})^{(k_{\rm \rho1}-k_{\rm \rho2})}~z > z_{\rm \rho0}
\end{cases}
\end{equation}

\begin{equation}
\begin{cases}
 L^* = L^*_{0} (1+z)^{k_{\rm l1}} ~~~~~~~~~~~~~~~~~~~~~~~ z < z_{\rm l0}\\
 L^* = L^*_{0} (1+z)^{k_{\rm l2}} (1+z_{\rm l0})^{(k_{\rm l1}-k_{\rm l2})}~~ z > z_{\rm l0}
\end{cases}
\end{equation}

where $\Phi^*_{0}$ and $L^*_{0}$ are the normalization and characteristic luminosity at $z=0$, $k_{\rm \rho1}$, $k_{\rm \rho2}$, $k_{\rm l1}$ and $k_{\rm l2}$ are the exponents for values lower and greater than $z_{\rm \rho0}$ and $z_{\rm l0}$ for $\Phi$ and $L$, respectively.
In this second fit, each point of the LF is associated to a redshift, corresponding to the median value of the galaxy population inside the individual redshift bin.
By using these values, we can trace a more accurate evolution of $\Phi^*$ and $L^*$. From this fit we obtain two evolution curves ($L^*$ and $\Phi^*$ vs $z$), which take into account the shape of the LF at each $z$. The priors used in this MCMC are given for the parameters that regulate the evolution of the LF, while $\alpha_S$ and $\sigma_S$ are fixed to 1.2 and 0.5 (found by \cite{gruppioni2013lf} at $z \sim 0.15$); and anchor $\Phi^*$ and $L^*$ to their value at $z\sim 0.7$ found with Herschel-only measurements. In particular, we fixed the $L^*$ and $\Phi^*$ at $z \sim 0.7$ to be the same found by Herschel.
The best-fit curve at each $z$-bin is showed in Figure \ref{fig:LF_models} as a dark-red solid line and shaded errors. %Compared to the individual fits, at $z<3$ the two curves are very similar, with the shape dominated by the presence of Herschel points; at $z \sim 3.25$ and $z \sim 4$ the evolutionary best-fit curve is characterized by slightly lower luminosities at the knee with respect to the individual fit, but they are still consistent within the errors. Finally, in the highest redshift bin, the purple curve is higher then the red one, but consistent within uncertainties. 
\par
\begin{figure}[]
\centering
{\includegraphics[width=.5\textwidth]{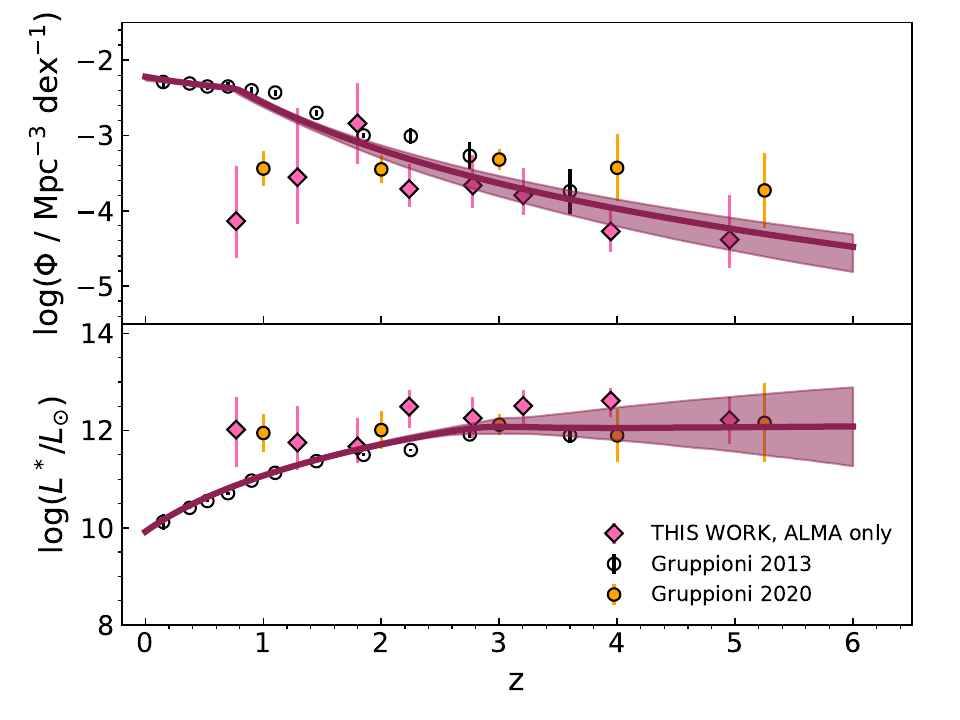}}
\caption{\small{Evolutionary trends of $\Phi^*$ (top panel) and $L^*$ (bottom panel) with redshift. The solid dark-red lines and shaded areas show the behaviours of $\Phi^*$ and $L^*$ at all the redshifts, while pink diamonds are the estimates obtained fitting each redshift bin individually in the ALMA only case. Finally, black empty circles and yellow points represent estimates for $L^*$ and $\Phi^*$ from \cite{gruppioni2013lf} and \cite{gruppioni2020alpine}, reported for comparison.}}
\label{fig:lstar_phistar}
\end{figure}

 The trend with redshift of $L^*$ and $\Phi^*$ is reported in Figure \ref{fig:lstar_phistar} and the results of the MCMC fit are reported in Tables \ref{tab:lf_mcmc_evol} and \ref{tab:lf_mcmc_evol_0}, representing the 16th and 84th percentiles errorbands and the best-fit values at $z=0$. The curve from the evolutionary MCMC and uncertainties are reported as dark-red line and shaded areas. The estimates of the individual fit are instead plotted for comparison as pink diamond for the ALMA-only case. From Figure \ref{fig:lstar_phistar} it is possible to notice that the values of $L^*$ and $\Phi^*$ estimated by fitting only the \a3 data points are slightly inconsistent with the results obtained instead using \a3 plus Herschel (at $z>2$). As mentioned before, this is due to the limited ability of ALMA to trace the dust peak emission at $z < 2$ and to the larger weight of Herschel data (containing more sources) in the combination.
The evolution of $\Phi$ changes from $z<z_{\rm \rho0} = 0.89^{+0.07}_{-0.14}$ to lower values, having an evolutionary trend of $\Phi^* \propto (1+z)^{-0.55}$ for $z<z_{\rm \rho0}$ and $\Phi^* \propto (1+z)^{-3.41}$ for $z>z_{\rho0}$. Overall, a decreasing trend is observed. This can be interpreted as a decreasing density of the bulk of star-forming galaxies at a given redshift. $L^*$ is showing two different trends with redshift below and above the break. In particular, for $z<z_{l0}=3.03^{+0.87}_{-0.73}$, the $L^*$ evolves as $(1+z)^{3.41}$ and as $L^* \propto (1+z)^{0.59}$ for $z>z_{\rm l0}$, thus becoming flatter. The evolutionary trend of $L^*$ can be ascribed to downsizing \citep{thomas2010down}, i.e., brighter (massive according to the $SFR-M$ relation) galaxies formed earlier than their fainter counterparts. A similar analysis has already been performed by \cite{gruppioni2013lf} using the \textit{Herschel} PEP/HerMES LF (black empty circles in Figure \ref{fig:lstar_phistar}), finding a consistently decreasing $\Phi^*$ and increasing $L^*$. However, their value of $z_{\rm l0}$ is lower than what we obtain ($z_{\rm l0} \sim 3$), being $z_{\rm l0} \sim 2$. The evolution is in a very good agreement with the Herschel only results (black empty points) and with the ALPINE estimates at $z>4$.

%\begin{table}[]
%\centering
%\renewcommand{\arraystretch}{1.5}
%\caption{Luminosities (L$^*$) and normalizations ($\Phi^*$), with errors, at knee, in the eight redshift bins, obtained through the MCMC analysis in the ALMA + \textit{Herschel} case.}
%\begin{tabular}{ccc}
%\hline
%\hline
%$z$       & log($L^*/L_{\odot})$    & log($\Phi/[\rm Mpc^{-3} \rm dex^{-1}$)  \\ \hline
%$0.5-1.0$ & $10.77_{-0.04}^{+0.04}$ & $-2.44_{-0.07}^{+0.08}$        \\
%$1.0-1.5$ & $11.41_{-0.04}^{+0.04}$ & $-2.95_{-0.07}^{+0.07}$        \\
%$1.5-2.0$ & $11.77_{-0.10}^{+0.09}$ & $-3.34_{-0.16}^{+0.17}$        \\
%$2.0-2.5$ & $11.56_{-0.07}^{+0.07}$ & $-2.87_{-0.14}^{+0.14}$        \\
%$2.5-3.0$ & $12.01_{-0.08}^{+0.08}$ & $-3.46_{-0.15}^{+0.16}$        \\
%$3.0-3.5$ & $12.48_{-0.35}^{+0.47}$ & $-3.68_{-0.37}^{+0.27}$        \\
%$3.5-4.5$ & $12.55_{-0.30}^{+0.38}$ & $-4.15_{-0.36}^{+0.30}$        \\
%$4.5-6.0$ & $12.23_{-0.48}^{+0.51}$ & $-4.29_{-0.61}^{+0.40}$        \\ \hline
%\end{tabular}
%\label{tab:lf_mcmc}
%\end{table}

\begin{table*}[]
\centering

\renewcommand{\arraystretch}{1.5}
\caption{Best fit parameters at the knee of the IR-LF.}
\begin{threeparttable}
\begin{tabular}{ccccccc}
\hline
\hline
$z$       & \begin{tabular}[c]{@{}c@{}}log($L^*/L_{\odot})$ \\ 16th\end{tabular} & \begin{tabular}[c]{@{}c@{}}log($L^*/L_{\odot})$ \\ 50th\end{tabular} & \begin{tabular}[c]{@{}c@{}}log($L^*/L_{\odot})$\\  84th\end{tabular} & \begin{tabular}[c]{@{}c@{}}log($\Phi^*/\rm Mpc^{-3} \rm dex^{-1}$)\\  16th\end{tabular} & \begin{tabular}[c]{@{}c@{}}log($\Phi^*/\rm Mpc^{-3} \rm dex^{-1}$)\\  50th\end{tabular} & \begin{tabular}[c]{@{}c@{}}log($\Phi^*/\rm Mpc^{-3} \rm dex^{-1}$)\\  84th\end{tabular} \\ \hline
$0.5-1.0$ & 10.93                                                                & 10.94                                                                & 10.95                                                                & -2.49                                                                      & -2.44                                                                & -2.40                                                                      \\
$1.0-1.5$ & 11.27                                                                & 11.29                                                                & 11.31                                                                & -2.84                                                                      & -2.77                                                                & -2.73                                                                      \\
$1.5-2.0$ & 11.58                                                                & 11.61                                                                & 11.66                                                                & -3.19                                                                      & -3.09                                                                & -3.04                                                                      \\
$2.0-2.5$ & 11.79                                                                & 11.84                                                                & 11.90                                                                & -3.44                                                                      & -3.32                                                                & -3.24                                                                      \\
$2.5-3.0$ & 11.93                                                                & 12.05                                                                & 12.15                                                                & -3.71                                                                      & -3.55                                                                & -3.46                                                                      \\
$3.0-3.5$ & 11.93                                                                & 12.07                                                                & 12.27                                                                & -3.90                                                                      & -3.71                                                                & -3.61                                                                       \\
$3.5-4.5$ & 11.81                                                                & 12.05                                                                & 12.47                                                                & -4.20                                                                      & -3.97                                                                & -3.84                                                                      \\
$4.5-6.0$ & 11.57                                                                & 12.07                                                                & 12.69                                                                & -4.53                                                                      & -4.24                                                                & -4.10  \\ \hline                                                                   
\end{tabular}
\begin{tablenotes}
   \small{ \item[*]{Luminosities (L$^*$) and normalizations ($\Phi^*$), with 16th, 50th and 84th percentiles, at knee, in the eight redshift bins, obtained through the MCMC analysis, in the ALMA+Herschel case, using the information from all the redshifts together.}}  
\end{tablenotes}
\end{threeparttable}
\label{tab:lf_mcmc_evol}
\end{table*}

\begin{table*}[]
\centering
\renewcommand{\arraystretch}{1.5}
\caption{Values at $z=0$ obtained from the MCMC evolutive fit.}
\begin{threeparttable}
\begin{tabular}{llllllllll}
\hline
\hline
log($L^*_{0}$)            & log$\Phi^*_{0}$             & $kl_{1}$               & $kl_{2}$            & $z_{l}$                & $k \rho_{1}$            & $k \rho_{2}$            & $z_{\rho}$             & $\alpha$ & $\sigma$ \\ \hline
$9.90_{-0.07}^{+0.07}$ & $-2.20_{-0.07}^{+0.07}$ & $3.41_{-0.16}^{+0.83}$ & $0.59_{-*}^{+*}$ & $3.03_{-0.73}^{+0.87}$ & $-0.55_{-0.16}^{+0.62}$ & $-3.41_{-0.67}^{+0.31}$ & $0.89_{-0.14}^{+0.07}$ & $1.2^f$  & $0.5^f$ \\ \hline
\end{tabular}
\begin{tablenotes}
   \small{ \item[*]{Values at $z=0$ obtained from the MCMC evolutive fit. $\alpha_{S}$ and $\sigma_{S}$ are fixed to the values found by \cite{gruppioni2013lf}.}}  
\end{tablenotes}
\end{threeparttable}

\label{tab:lf_mcmc_evol_0}
\end{table*}

\section{Dust-obscured star formation rate density}\label{sec:sfrd}

\begin{figure*}[]
\centering
{\includegraphics[width=.9\textwidth]{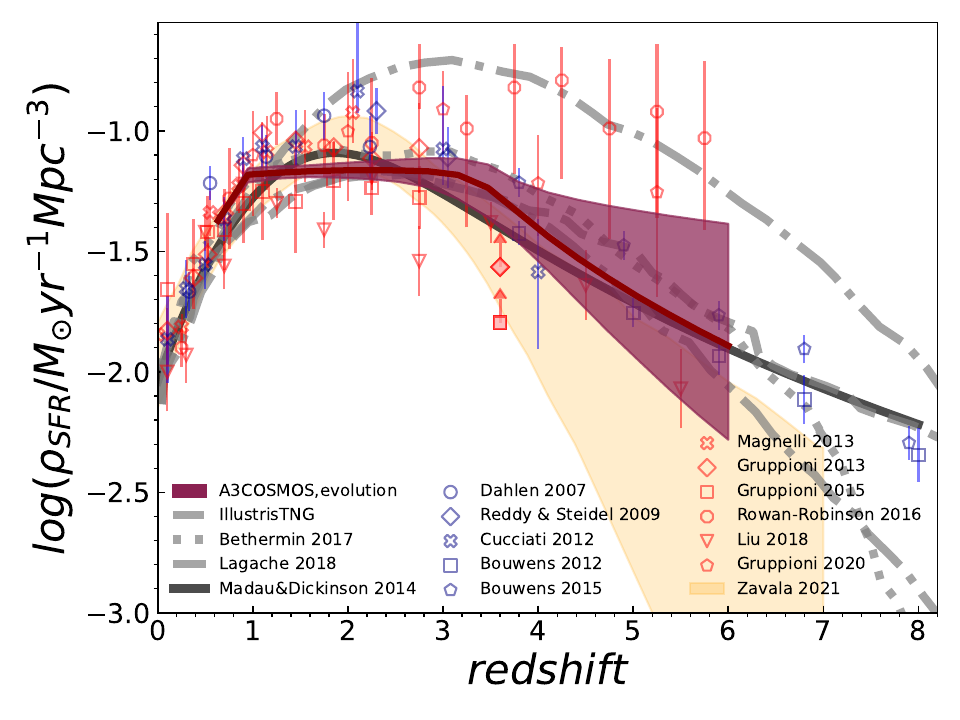}}
\caption{\small{Comoving star formation rate density evolution with $z$. The red diamonds are obtained through the integration of the LF best-fit in each $z$ bin, while the dark-red shaded area is obtained by integrating the evolutionary LF from $z=$ 0.5 to 6. Comparison curves from literature are reported as grey dash-dotted (pessimistic and optimistic cases, \textcolor{blue}{Lagache et al. 2018}), dashed (IllustrisTNG), dotted (\textcolor{blue}{Bethermin et al. 2017}) and black solid (\textcolor{blue}{Madau \& Dickinson 2014}) lines. The red empty points with different symbols represent different IR-mm estimates of the SFRD from \cite{gruppioni2013lf, magnelli2013ir, gruppioni2015sfrd,rowanrobinson2016sfrd,liu2018sfrd,gruppioni2020alpine} and the orange shaded area is the dust-obscured SFRD by \cite{zavala2021mora}, using the MORA survey. SFRD from the UV is plotted as blue empty points with different markers \citep{Dahlen2007sfrd,reddy2009sfrd,cucciati2012sfrd,bouwens2012sfrd, bouwens2015sfrf} \\.}}
\label{fig:sfrd}
\end{figure*}

\begin{table}[]
\centering
\renewcommand{\arraystretch}{1.5}
\caption{SFRD obtained by integrating the LF best-fit in our eight redshift bins.}
\begin{threeparttable}

\begin{tabular}{cccc}
\hline
\hline
$z$       & \begin{tabular}[c]{@{}c@{}}log$\rho_{\rm SFR}$ \\ {[}$\rm M_{\odot} \rm yr^{-1} \rm Mpc^{-3}${]}\\ 50th\end{tabular} & \begin{tabular}[c]{@{}c@{}}log$\rho_{\rm SFR}$ \\ \\ 16th\end{tabular} & \begin{tabular}[c]{@{}c@{}}log$\rho_{\rm SFR}$ \\ \\ 84th\end{tabular} \\ \hline
$0.5-1.0$ & $-1.25$                                                                                                 & $-1.27$                                                                                                 & $-1.23$                                                                                                 \\
$1.0-1.5$ & $-1.17$                                                                                                 & $-1.20$                                                                                                 & $-1.15$                                                                                                 \\
$1.5-2.0$ & $-1.16$                                                                                                 & $-1.19$                                                                                                 & $-1.14$                                                                                                 \\
$2.0-2.5$ & $-1.16$                                                                                                 & $-1.20$                                                                                                 & $-1.13$                                                                                                 \\
$2.5-3.0$ & $-1.17$                                                                                                 & $-1.23$                                                                                                 & $-1.12$                                                                                                 \\
$3.0-3.5$ & $-1.19$                                                                                                 & $-1.29$                                                                                                 & $-1.12$                                                                                                 \\
$3.5-4.5$ & $-1.40$                                                                                                 & $-1.53$                                                                                                 & $-1.24$                                                                                                 \\
$4.5-6.0$ & $-1.67$                                                                                                 & $-1.92$                                                                                                 & $-1.32$    \\ \hline                                                                                            
\end{tabular}

\begin{tablenotes}
   \small{ \item[*]{The second column is the median value, while third and fourth columns show the lower and upper 16th boundaries.}}  
\end{tablenotes}
\end{threeparttable}
\label{tab:sfrd_values}
\end{table}

Finally, using the best-fit luminosity function in each redshift bin, we derive the comoving SFRD. In particular, we first obtain the IR luminosity density ($\rho_{\rm IR}$) by integrating the IR LF in each $z$ bin, from $log(L_{\rm IR}) = 8$ (to be consistent with others IR-based SFRDs) to $log(L_{\rm IR}) = 14$ %(we divided the redshift array into 20 equally spaced bins of redshift, but we tabulate our results in the eight bins covered by our LF)
:
\begin{equation}
    \rho_{\rm IR}(z)=\int_{8}^{14} \Phi (logL_{\rm IR,z})L_{\rm IR}~dlogL_{\rm IR}
\end{equation}

Then, we can convert the IR luminosity density into a SFRD by applying the \cite{kennicutt1998sfr} relation to convert $L_{\rm IR}$ to SFR, assuming a \cite{chabrier2003imf} IMF. As described in Section \ref{mcmc}, in addition to the individual fit, we performed an MCMC fitting taking the information from all the redshift bins together to derive the redshift evolution of $L^*$ and $\Phi^*$. In this way, we obtained the LF at each redshift from $z \sim 0.5$ to $z \sim 6$ and we integrated the LF[$L^*(z)$,$\Phi^*(z)$)] in the full redshift range (for clarity we report the results of the fit for the eight bins of our IR LF). The result is an evolution curve of the SFRD with the redshift, shown in Figure \ref{fig:sfrd}. The values of the SFRD and errorbands, are reported in Table \ref{tab:sfrd_values}. The dark-red solid lines, with the shaded area, represent the lower and upper boundaries of the curve, derived as the errors on the integration of the LF at each redshift. In order to compare our values with those from literature, we overplot previous results on the SFRD from \cite{madau2014sfrd,bethermin2017sfrd,lagache2018sfrd} and from the IllustrisTNG simulation \citep{pillepich2018illustris}. Estimates of the dust-obscured SFRD from other IR-mm works \citep[][]{gruppioni2013lf, magnelli2013ir, gruppioni2015sfrd,rowanrobinson2016sfrd,liu2018sfrd,gruppioni2020alpine} are reported as red empty points with different markers. Blue empty points represent the SFRD from UV works, corrected from dust attenuation, by \cite{Dahlen2007sfrd,reddy2009sfrd,cucciati2012sfrd,bouwens2012sfrd,bouwens2015sfrf}. 
\par Thanks to the usage of Herschel data points, combined with the ALMA ones, in the LF fit, we were able to derive accurate estimates of the dust-obscured SFRD also at $z < 3$. We found the SFRD computed at $z \sim 0.5-1.0$ follow the rise described by the \cite{madau2014sfrd} black curve and is consistent with the IR and UV estimates from previous works. Although our points up to $z \sim 2$ have a lower normalization with respect to those from \cite{gruppioni2013lf} and \cite{magnelli2013ir}, this discrepancy may be explained by the different type of fit performed which results in a different extrapolation to the faint-end. From $z \sim 2$ to $z \sim 3.5$ the SFRD is following a flat trend, being above the \cite{madau2014sfrd}. This flattening of the SFRD is compatible with predictions from models that envisage the early formation of massive spheroids \citep[see ][]{2003ApJ...596..734C}.
As mentioned before, at $z>3$ our estimates are derived using ALMA alone (last three redshift bins). The observed trend is a decrease of the dust-obscured SFRD up to $z \sim 6$, even though the limited number of sources does not allow us to strictly constrain it at those redshifts. We found the $z>3$ dust-obscured SFRD to be consistent with the dust-corrected UV estimates and with the \cite{madau2014sfrd} one and, in the lower boundary, with the dust-obscured SFRD derived by \cite{zavala2021mora}, although with a $\sim 0.5$ dex difference between the mean values. The upper boundary is consistent with \cite{gruppioni2020alpine}, pointing towards a possibly underestimated SFRD at $z>3$. However, this underestimation seems not to be as extreme as suggested by \cite{rowanrobinson2016sfrd} and \cite{gruppioni2020alpine}.

\label{fig:LF_comparison}
%\end{figure*}

%\subsection{The AGN-corrected SFRD}
%Finally, as described in Section \ref{sec:sfrd}, we derive the SFRD AGN-corrected SFRD directly from the SF-related IR luminosity function. As for the LF, the differences between the two SFRD are not significant. We find a slight decrease in normalization for the value of SFRD in the last redshift bin ($4.5 < z < 6.0$). Thus, we can argue that the presence of an AGN in the host galaxy may affect the individual L$_{IR}$ estimates, but it has less relevance when deriving the binned statistical properties of the sample (i.e., LF and SFRD).

%\begin{figure}[]
%\centering
%{\includegraphics[width=.45\textwidth]{SF_only/sfrd_comparison_tot_sf.pdf}}
%\caption{\small{Comoving star formation rate density evolution with $z$. The blue boxes are obtained through the integration of the total IR LF best-fit in each $z$ bin, while the red points and boxes are obtained through the integration of the SF-only IR LF best-fit.\\}}
%\label{fig:sfrd_comparison}
%\end{figure}
%%%%%%%%%%%%%%%%%%%%%%%%%%%%%%%%%%%
%
% Discussion
%
%%%%%%%%%%%%%%%%%%%%%%%%%%%%%%%%%%%%

\section{Summary and Conclusions}\label{discussion}

Owing to the COSMOS2020 + ALMA photometric coverage, we were able to perform SED fitting using the \texttt{CIGALE} code to constrain the SFR, $L_{\rm IR}$ and $M_{\rm \star}$ of the full ALMA sample. Applying the \cite{A&B1980sample} method we were able to deal with the overlapping regions between several pointings, to evaluate correctly the depth of those regions. Moreover, we derived an SED-based method to homogenize the pointings at different observing wavelengths converting the RMS noise into that of the reference wavelength. By removing pointings without a detected target and by correcting for clustering, we were able to turn \a3 into an untargeted, “blind-like” survey. We thus derived the total infrared luminosity function in the $8-1000 \mu$m band using the $1/V_{MAX}$ \citep[][]{Schmidt1970vmax} method and using the already available data from \textit{Herschel}. We estimated the LF from $z \sim 0.5$ to $z \sim 6$ and performed a MCMC simulation to fit the LF data, including the possibility for the characteristic parameters ($\Phi^*$ and $L^*$) to evolve with redshift. By integrating the LF we derived the infrared luminosity density and, thus, the dust-obscured star formation rate density up to $z \sim 6$.

\par We summarize the conclusions of this work as follows:
\begin{itemize}
 \item We find the \a3 sample to be characterized by galaxies that are massive (log($M_{\star}/\rm M_{\odot}) = 10-12$) and bright in the IR ($8-1000 \mu$m) band, with log($L_{\rm IR}/\rm L_{\odot}) = 11-13.5$. Converting $L_{\rm IR}$ into SFR, we obtained values typical of normal star forming (SFR $\sim 1-100$ $\rm M_{\odot}$yr$^{-1}$) and starbursting galaxies (SFR $\sim 100 - 1000$ $\rm M_{\odot}$yr$^{-1}$).
\\
 \item We find our LF to be in a good agreement with existing literature (in particular at $z > 1$), though pushing the SFRD to $z\sim 6$ with unprecedented statistics. 
\\
 \item Our MCMC analysis suggests a joint redshift-decreasing number density and a redshift-increasing IR luminosity for ALMA selected star-forming galaxies. This result is consistent with these galaxies being less frequent and more luminous (i.e., massive) going towards higher redshift.
\\
 \item Our estimates of the dust-obscured SFRD are consistent with those from other IR surveys and with the UV dust-corrected estimates. Also, we found a broader peak in the SFRD, with a smooth decrease at $z>4$, which suggests a significant contribution from the obscured SFRD at high redshift. Furthermore, a contribution to the SFRD, particularly at high redshifts, could originate from HST-dark sources \citep[][]{franco2018dark,wang2019hst,talia2021dark,behiri2023hstdark}, which are not a priori included in a prior sample, with the exception of the 6 added sources.
\end{itemize}

In conclusion, our study of the physical (i.e., stellar mass, dust luminosity, star formation rate) and statistical (LF and SFRD) properties of the ALMA sample of galaxies in the COSMOS field resulted in the detection and characterisation of a starforming and starbursting dominated population, $\sim 40\%$ of which is likely hosting an AGN. The \ad sample is also found to evolve both in luminosity ($L^*$) and density ($\Phi^*$), and to significantly contribute to the total (IR+UV) SFRD also at $z>3$. In order to improve our knowledge and to put tighter constraints to the evolution and formation of galaxies at higher $z$, more statistics is however needed. Being constantly updated, the \a3 catalog will represent a key survey to reach newer results with a better statistics. The future COSMOS-Web survey \citep[][]{casey2022webb} will also play an important role on covering the COSMOS field with JWST, allowing to obtain a better photometric coverage and, thus, allowing the physical properties of galaxies to be better constrained.
%show an evolutionary trend in its characteristic luminosity (L$^*$) and normalization ($\Phi^*$) with redshift and significantly contributes to the total SFRD at $z>3$.

\section*{Acknowledgements}
ID acknowledges support from INAF Minigrant "Harnessing the power of VLBA towards a census of AGN and star formation at high redshift". LB acknowledges the support from grant PRIN MIUR $2017 - 20173ML3WW$\_$001$. ES acknowledges funding from the European Research Council (ERC) under the European Union’s Horizon 2020 research and innovation programme (grant agreement No. 694343). This work was supported by NAOJ ALMA Scientific Research Grant Code 2021-19A (HA). SG acknowledges financial support from the Villum Young Investigator grant 37440 and 13160 and the Cosmic Dawn Center (DAWN), funded by the Danish National Research Foundation (DNRF) under grant No. 140. AT acknowledges G. Mazzolari and A. della Croce for all the useful discussions and N. Borghi, L. Leuzzi and C. Perfetti for all the support during the writing of this article. 

\bibliographystyle{aa}
\bibliography{1biblio}

\begin{appendix}
\section{SED fitting input modules}\label{sec.foo}
%\section{SED fitting input modules}
In this Appendix, we discuss the main details of the input modules we selected for the SED fitting with CIGALE.
\subsection{Stellar population and star formation history}
Among the CIGALE options, we select the \texttt{bc03} module, i.e. the stellar population synthesis model by \cite{bruzual2003ssp}, to build the SED stellar component.%, coupled with the \cite{chabrier2003imf} initial mass function (IMF).
\par To parametrise the star formation history (SFH) we used the \texttt{sfhdelayed} module. The SFR initially increases up to a certain time defined as $t=\tau$ (with $\tau$ being the $e$-folding time of the main stellar population model) and then decreases, as described by the following analytical form:
\begin{equation}
    \rm SFR(t) \it \propto \frac{t}{\tau^2} \times exp(-t/\tau)~~for~0 \leq t \leq t_0
\end{equation}
with t$_0$ being the age of the onset of star formation. An optional burst representing the latest episode of star formation can be added \citep[][]{malek2018sed}. However, in this work, we choose to represent the SFH with a nearly constant shape, selecting 10.000 Myr as value for the $e$-folding time of the late starburst population model.

\subsection{Dust component}
As already explained in Section \ref{sec:intro}, the emission of DSFGs cannot be entirely reproduced without taking into account for the presence of dust in the galaxy. For this reason, we included in the SED fitting the modules representing dust attenuation and re-emission. 
\par We selected the \texttt{dustatt\_modified\_CF00} \citep[][]{charlot2000dustatt} template to model the attenuation by dust. The \texttt{dustatt\_modified\_CF00} assumes the presence of both diffuse interstellar medium (ISM) and birth cloud (BC) in the surroundings of stars. In particular, both young and old stellar populations' light is affected by the ISM attenuation, whereas the BCs attenuate only the emission from young stars, as they recently formed inside the clouds. Considering these two different but coexisting situations, two attenuation laws are computed, with different slopes for ISM and BCs. The $V$-band attenuation is then derived as the ratio between the ISM and the total (i.e., ISM + BCs) attenuation.
\par To consistently model the dust emission, we use the \texttt{dl2014} \citep[][]{draine2014dustem} template. Indeed, the \texttt{dl2014} module is based on the same ISM and BCs environment used for the dust attenuation. This module takes the PAH mass fraction and the minimum radiation field as main input parameters.

\subsection{AGN component}
\begin{figure}[]
\centering
{\includegraphics[width=.4\textwidth]{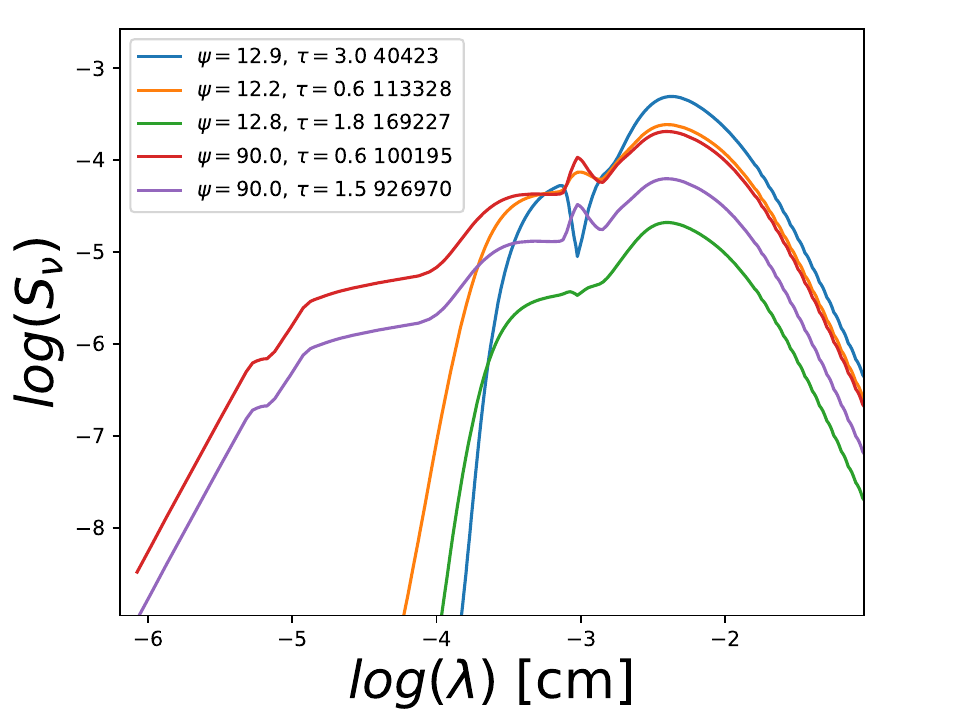}}
\caption{\small{Examples of AGN templates adopted in the SED fitting, with different values of optical depth at 9.7 $\mu$m ($\tau$) and the angle between equatorial axis and line of sight ($\Psi$). \\}}
\label{fig:cigale_agn}
\end{figure}

Finally, we use the \texttt{fritz2006} module \citep[][]{fritz2006agnmodel,feltre2012agnmodel} to model the AGN component in the SED. The AGN emission is described with a radiative trasfer model which takes into account a primary emission coming from the engine (i.e. the accretion disc), a scattered emission produced by dust and a thermal component of the dust emission. %The several free parameters permit to have a large possibility of AGN configurations. 
In particular, the main input parameters are the following: the ratio between the inner and outer radii of the dusty torus; the equatorial optical depth at 9.7 $\mu$m, the dust density distribution described by two parameters ($\propto r^{\beta} e^{-\gamma |cos\theta|}$); the opening angle of the dusty torus; the angle $\Psi$ between the equatorial axis and the line of sight and the AGN fraction ($f_{AGN}$), defined as the ratio between the AGN IR luminosity and the total IR luminosity (i.e., AGN + SF) in the same bands. In Figure \ref{fig:cigale_agn}, we show some AGN models with different $\tau$ and $\Psi$ obtained in the SED fitting. It is possible to see how different combinations of these parameters give rise to different shapes and types of AGN (i.e., Type 1 and Type 2).
\par In table \ref{tab:cigale} we report a summary of all the input parameters used for SED fitting with \texttt{CIGALE}.
\begin{table*}[]
\centering
\renewcommand{\arraystretch}{1.5}
\caption{\texttt{CIGALE} parameters used for the SED fitting.}

\begin{threeparttable}
\label{tab:cigale}
\begin{tabular}{ccc}
\hline
\hline
\multicolumn{1}{l}{}                                                                                                                               & \multicolumn{1}{l}{}                                                                                                                                                                 & \multicolumn{1}{l}{}                                                                                                                                                                                                                                                                                                             \\
\textbf{Parameter}                                                                                                                                 & \textbf{Values}                                                                                                                                                                      & \textbf{Description}                                                                                                                                                                                                                                                                                                             \\
\multicolumn{1}{l}{}                                                                                                                               & \multicolumn{1}{l}{}                                                                                                                                                                 & \multicolumn{1}{l}{}                                                                                                                                                                                                                                                                                                             \\ \hline
\begin{tabular}[c]{@{}c@{}}SFH (\texttt{sfhdelayed})\\ $\tau_{main}$\\ age$_{main}$\\ $\tau_{burst}$\\ age$_{burst}$\\ f$_{burst}$\end{tabular}             & \begin{tabular}[c]{@{}c@{}}\\1.0, 2.0, 5.0, 7.0 {[}Gyr{]}\\ 1.0, 3.0, 5.0, 8.0, 11.0, 12.0 {[}Gyr{]}\\ 10.0 {[}Gyr{]}\\ 0.001, 0.01, 0.1, 0.3 {[}Gyr{]}\\ 0.0, 0.15, 0.30\end{tabular} & \begin{tabular}[c]{@{}c@{}}\\$e$-folding time of the main stellar population\\ Age of the main stellar population in the galaxy\\ $e$-folding time of the late starburst population\\ Age of the late burst\\ Mass fraction of the late burst population\end{tabular}                                                              \\ \hline
\begin{tabular}[c]{@{}c@{}}Stellar component (\texttt{bc03})\\ IMF\\ Z\\ Separation age\end{tabular}                                                        & \begin{tabular}[c]{@{}c@{}}\\1\\ 0.02\\ 10 {[}Myr{]}\end{tabular}                                                                                                                      & \begin{tabular}[c]{@{}c@{}}\\ \\Initial mass function: 1 (Chabrier 2003)\\ Metallicity \\ Age of the separation between \\ the young and the old star populations\end{tabular}                                                                                                                                                         \\ \hline
\begin{tabular}[c]{@{}c@{}}Dust attenuation \\ (\texttt{dustatt\_modified\_cf00})\\ A$^{ISM}_V$\\ $\mu$\\ slope ISM\\ slope BC\end{tabular} & \begin{tabular}[c]{@{}c@{}}\\ \\0.3, 1.7, 2.8, 3.3\\ 0.3, 0.5, 0.8, 1.0\\ -0.7\\ -0.7\end{tabular}                                                                                        & \begin{tabular}[c]{@{}c@{}}\\ \\V-band attenuation in the interstellar medium\\ Ratio of the BC-to-ISM attenuation\\ Power law slope of the attenuation in the ISM\\ Power law slope of the attenuation in the BC\end{tabular}                                                                                                        \\ \hline
\begin{tabular}[c]{@{}c@{}}Dust emission (\texttt{dl2014})\\ q$_{PAH}$\\ U$_{min}$\\ $\alpha$\\ $\gamma$\end{tabular}                                       & \begin{tabular}[c]{@{}c@{}}\\0.47, 2.5, 3.9\\ 0.1, 1.0 5.0, 10.0, 25.0, 40.0\\ 1.0, 2.0, 3.0\\ 0.0, 0.02\end{tabular}                                                                                          & \begin{tabular}[c]{@{}c@{}}\\Mass fraction of PAH\\ Minimum radiation field\\ Dust emission power law slope\\ Illuminated fraction\end{tabular}                                                                                                                                                                                    \\ \hline
\begin{tabular}[c]{@{}c@{}}AGN component (\texttt{fritz2006})\\ r$_{ratio}$\\ $\tau$\\ $\beta$\\ $\gamma$\\ Opening angle\\ $\Psi$\\ $f_{AGN}$\end{tabular} & \begin{tabular}[c]{@{}c@{}}\\60.0\\ 0.6, 1.0, 3.0\\ -0.5\\ 0.0\\ 100.0\\ 0.001, 30.100, 89.99\\ 0.0, 0.1, 0.15, 0.25, 0.50\end{tabular}                                                & \begin{tabular}[c]{@{}c@{}}\\Ratio of the maximum to minimum radii of the dusty torus\\ Equatorial optical depth at 9.7 $\mu$m\\ Radial dust distribution within the torus\\ Angular dust distribution within the torus\\ Full opening angle of the dusty torus\\ Angle between equatorial axis and line of sight\\ AGN fraction\end{tabular} \\ \\ \hline \hline \\
\end{tabular}

\begin{tablenotes}
   \small{ \item[*]{The first column reports the name of the templates as well as each individual parameter. In the second column the parameters are reported and in the third column the descriptions of the parameters are given.}}  
\end{tablenotes}
\end{threeparttable}

\end{table*}

%\section{Possible HST-dark}\label{hstdark}
%\textbf{
%We show, in this Appendix, some example of the potential HST-dark sources, with their SEDs and the optical/NIR + ALMA cutouts.
%}
\end{appendix}

\end{document}